\documentclass[%
aip,
amsmath,amssymb,
reprint,%
]{revtex4-2}

\usepackage[pdftex]{graphicx}

\usepackage{dcolumn}
\usepackage{bm}

\usepackage{bm}        
\usepackage{amsfonts}  
\usepackage{amsmath}   
\usepackage{amssymb}   
\usepackage{subfig}
\usepackage{subfloat}  
\usepackage{float}
\usepackage{bm}

\usepackage[utf8]{inputenc}
\usepackage[T1]{fontenc}
\usepackage{mathptmx}
\usepackage{etoolbox}
\usepackage{multirow}

\usepackage{comment}

\usepackage{color,soul}
\usepackage{booktabs}
\usepackage[table,xcdraw]{xcolor}
\usepackage{algorithm,algpseudocode}
\usepackage{mathtools}
\usepackage{siunitx}
\usepackage[misc]{ifsym}

\newcommand{\grad}{\mbox{$\nabla$}}
\newcommand{\divg}{\mbox{$\nabla\cdot\,$}}

\usepackage[colorlinks=true,
linkcolor=cyan,
citecolor=cyan,
urlcolor=cyan,
]{hyperref}%

\usepackage{lineno}

\usepackage{xcolor}

\makeatletter
\def\@email#1#2{%
	\endgroup
	\patchcmd{\titleblock@produce}
	{\frontmatter@RRAPformat}
	{\frontmatter@RRAPformat{\produce@RRAP{*#1\href{mailto:#2}{#2}}}\frontmatter@RRAPformat}
	{}{}
}%
\makeatother
\begin{document}
	
	\preprint{AIP/123-QED}
	
	\title{Evolution of waves in liquid films on moving substrates}
	
	\author{Tsvetelina Ivanova}
	\affiliation{
		von Karman Institute for Fluid Dynamics, Waterloosesteenweg 72, Sint-Genesius-Rode, Belgium
	}
	\author{Fabio Pino}
	\affiliation{
		von Karman Institute for Fluid Dynamics, Waterloosesteenweg 72, Sint-Genesius-Rode, Belgium
	}
	\author{Benoit Scheid}
	\affiliation{
		TIPs lab, Université Libre de Bruxelles, Belgium
	}
	\author{Miguel A. Mendez}
	\email{mendez@vki.ac.be}
	\affiliation{
		von Karman Institute for Fluid Dynamics, Waterloosesteenweg 72, Sint-Genesius-Rode, Belgium
	}

	\date{\today}
	
	\begin{abstract}
		Accurate and computationally accessible models of liquid film flows allow for optimizing coating processes such as hot-dip galvanization and vertical slot-die coating. This paper extends the classic three-dimensional integral boundary layer (IBL) model for falling liquid films (FF) to account for a moving substrate (MS). We analyze the stability of the liquid films on vertically moving substrates in a linear and in a nonlinear setting. In the linear analysis, we derive the dispersion relation and the temporal growth rates of an infinitesimal disturbance using normal modes and linearized governing equations. In the nonlinear analysis, we consider disturbances of finite size and numerically compute their evolution using the set of nonlinear equations in which surface tension has been removed. We present the region of (linear) stability of both FF and MS configurations, and we place the operating conditions of an industrial galvanizing line in these maps. A wide range of flow conditions was analyzed and shown to be stable according to linear and nonlinear stability analyses. Moreover, the nonlinear analysis, carried out in the absence of surface tension, reveals a nonlinear stabilizing mechanism for the interface dynamics of a liquid film dragged by an upward-moving substrate.
	\end{abstract}
	
	\maketitle

	\section{Introduction} 
	\label{sec:Intro}
	The evolution of waves in liquid films plays a fundamental role in many coating processes. Their occurrence delimits the range of operating conditions and influences the quality of final products in the coating industry. 
	
	In falling liquid films, waves naturally develop and evolve through various patterns due to a fascinating interaction between inertial, viscous, gravitational, and capillary forces\cite{falling_liquid_films}. 
	One approach for investigating these waves is by integral boundary layer (IBL) models (see \citet{shkadov}). These models of lower-dimensionality proved to be useful in simulating industrial processes such as hot dip galvanization\cite{mendez_jfm}, which are not yet accessible by direct high fidelity simulations because of the prohibitive computational cost (see \citet{Aniszewski2020} and \citet{BarreiroVillaverde2021}). In these models, the dynamics of the liquid film flow is described in terms of film thickness and streamwise and spanwise flow rates, as opposed to the Navier-Stokes equations where the film thickness, velocity and pressure fields must be computed.  
	
	The literature on the integral modeling of falling liquid films is vast, with pioneering two-dimensional formulation proposed by \citet{kapitza_2D, shkadov} and later extended to three-dimensional models by \citet{demekhin_shkadov} (see also \citet{demekhin_2007, demekhin_2010}). 
	Improvements over the classic self-similar formalism have been proposed by \citet{RuyerQuil1998,RuyerQuil2000}, who solved the inconsistency in the prediction of the stability threshold by using the method of weighted residuals (see also \citet{SCHEID2006} for a three-dimensional extension). An extensive review of the modeling of falling liquid film is proposed by \citet{ruyer-quil} and by \citet{falling_liquid_films}. 
	
	Integral models enable analytical insight into the flow's stability ranges and enable computationally inexpensive simulations of their nonlinear dynamics. The numerical advantages of using integral models for falling films are also illustrated in \citet{scheid_3D_flow_structures} and \citet{scheid_wavemaker}. Recently, integral models have been extended by \citet{mendez_jfm} to the problem of liquid films evolving on a moving substrate in the presence of pressure gradient and shear stress at the interface. This configuration is encountered in the jet wiping process in hot-dip galvanization (see \citet{Buchlin},\citet{Gosset2019} \citet{Mendez2019}). Integral models allowed for analyzing the liquid film response to various disturbances in the process (see also \citet{Hocking2010} and \citet{BarreiroVillaverde2021}). Although the configuration was limited to 2-D models, the simulations suggest that thin films are more stable on an upward-moving substrate than on a fixed one. 
	
	This work aims to investigate the reasons for this difference and analyze how the kinematic condition at the wall influences the stability of the liquid interface. Moreover, we extend the 2-D models in \citet{mendez_jfm} to a 3-D configuration similar to the model by \citet{demekhin_2007}, here generalized to account for the substrate motion. The formulation of the model includes the presence of shear stress and pressure gradient exerted at the interface by an external flow, even though these are disregarded in our analysis.
	We first use the derived model to perform a classic linear stability analysis via normal modes and compare the results with the well-known case of falling liquid films. We then investigate the flow stability numerically in a nonlinear framework by studying how disturbances of finite size evolve in the film according to the (nonlinear) set of equations in the absence of surface tension. Surprisingly, the results show that nonlinearities have a stabilizing effect since conditions that are linearly unstable are nonlinearly stable.
	
	The rest of the article is structured as follows. Section \ref{sec2} reviews the reference quantities used to scale the dynamics of a liquid film in both the case of fixed or moving substrate. Section \ref{sec3} introduces the integral models while section \ref{sec4} presents the linear stability analysis for both cases. Section \ref{sec5} describes the numerical methods implemented in an in-house finite volume solver for the nonlinear partial differential equations (PDEs) derived in section \ref{sec3}. 
	The numerical implementations for a fixed and a moving substrate are validated in \ref{sec5p1} and \ref{sec5p2}, respectively. Section \ref{sec5p3} introduces the test cases analyzed in this work. The results of the stability analysis are collected in Section \ref{sec6} for both the linear (\ref{sec6p1}) and the nonlinear (\ref{sec6p2}) analysis. Conclusions and perspectives are in Section \ref{sec7}.
	
	\section{Scaling Laws} \label{sec2}
	
\begin{figure}[htbp]
		\centering
		\includegraphics[width=0.94\linewidth]{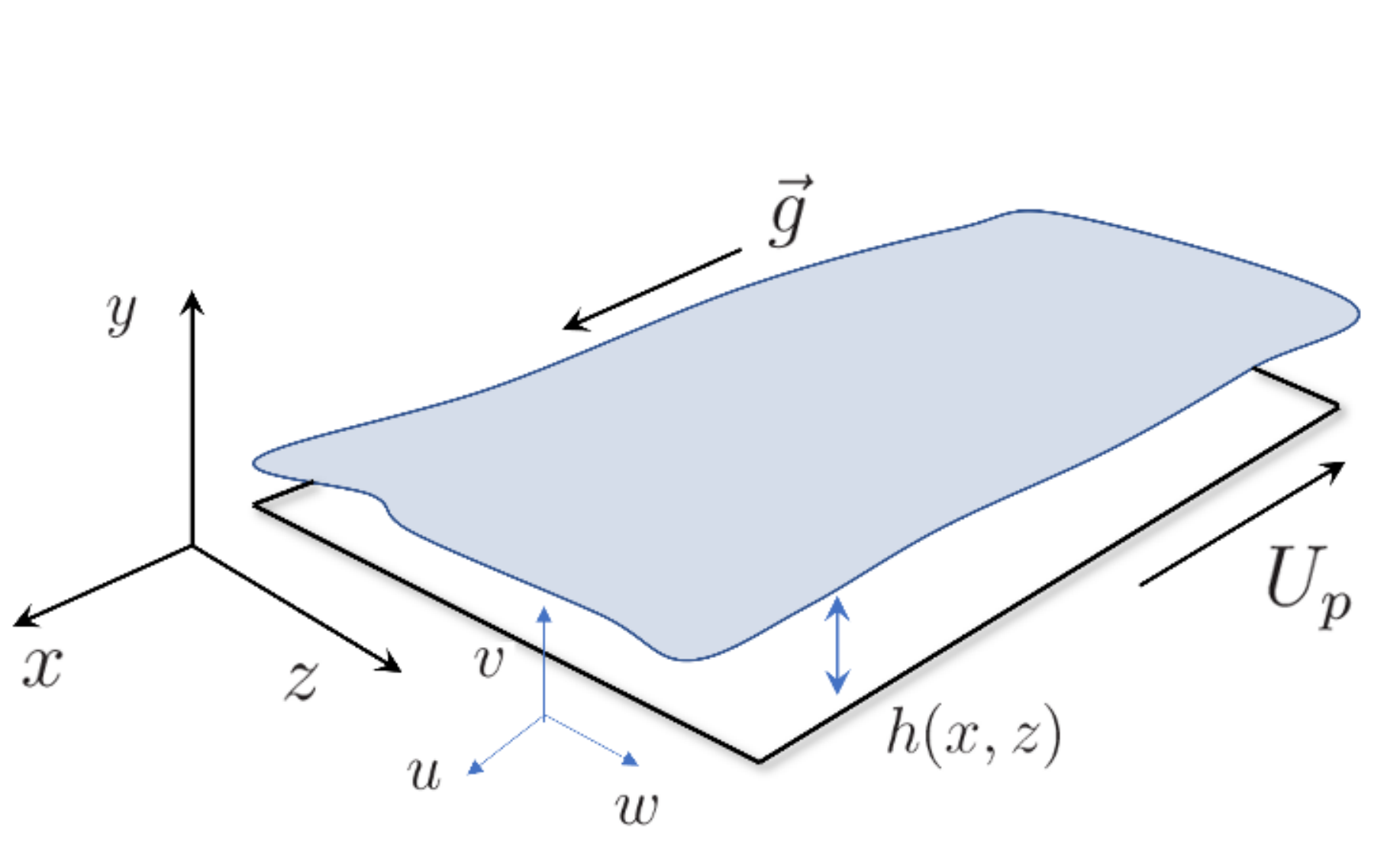}
		\caption{Sketch of the flow domain for a liquid film on a vertically moving substrate. Gravity is in the opposite direction of the substrate motion. }
		\label{fig:jw3d}
	\end{figure}

	The configuration of interest is three-dimensional and it is illustrated in Fig. \ref{fig:jw3d}. The liquid is assumed to be incompressible with kinematic viscosity $\nu$, density $\rho$, 
	dynamic viscosity $\mu=\rho\nu$, and surface tension $\sigma$. As illustrated in the figure, we consider gravity directed towards $x>0$, we set $y$ orthogonal to the substrate and $z$ in the spanwise direction. The pressure in the liquid is denoted by $p$, and the three velocity components are denoted as $\vec{u}=(u,v,w)$, oriented as shown in the figure. The flow is bounded
	by the substrate at $y=0$, and the dynamic liquid interface is at $y=h(x,z,t)$ ($h$ is also referred to as film thickness).
	
	In the classic falling film problem, herein denoted as `FF', the substrate is fixed. In the moving substrate problem, here denoted as `MS', it is moving at a velocity $U_p$ against gravity. The FF and MS problems are governed by different scaling laws.
	
	In the FF problem, one imposes the thickness (or the flow rate) of the film and sets the scales for the velocity from the viscous-gravity balance in steady conditions. Denoting as $h_N$ the (given) thickness in steady state conditions, and using square brackets to denote reference quantities such that $\hat{a}=a/[a]$ is the scaling of $a$ with respect to the reference $[a]$, for a falling film one has (see \citet{falling_liquid_films}):
	\begin{equation}
		\label{Scale_FF}
		[h]=h_N, \,\,\, \,\,\,[u]=\frac{g [h]^2}{\nu},  \,\,\, \,\,\, [q]=[u][h]=\frac{g h^3_N}{\nu}\,.
	\end{equation}
	Accordingly, the Reynolds number in the FF problem is defined as $Re=[q]/\nu=g h^3_N/\nu^2$.
	
	In the MS problem, one imposes the velocity of the substrate and it is thus natural to set $[u]=U_p$. The scale for the film thickness can also be computed from the viscous-gravity balance in steady state conditions, leading to:
	\begin{equation}
		\label{Scale_MS}
		[u]=U_p, \,\,\, \,\,\,[h]=\sqrt{\frac{\nu U_p}{g}} \,\,\, \,\,\, [q]=[u][h]=\sqrt{\frac{\nu U_p^3}{g}}\,.
	\end{equation}
	Accordingly, the Reynolds number in the MS problem is defined as $\text{Re}=[q]/\nu=\sqrt{U^3_p/g \nu}$.
	
	In both problems, it is convenient to scale the streamwise direction $x$ such that $\varepsilon =[h]/[x]\ll 1$ and the capillary forces ($\sim \sigma [h]/[x]^3$) balance the gravitational ones ($\sim \rho g$). This is known as Shkadov's scaling, (\citet{skhadov,falling_liquid_films}) and leads to
	\begin{subequations}
		\label{Epsilon}
		\begin{eqnarray} 
			\varepsilon&=&\Biggl(\frac{[h]^2 \rho g}{\sigma} \Biggr)^{1/3}=\text{We} ^{-1/3} \quad \mbox{ (in FF)},\\
			\varepsilon&=&\Biggl(\frac{[h]^2 \rho g}{\sigma} \Biggr)^{1/3}=\text{Ca} ^{1/3}\qquad \mbox{(in MS)}\,,
		\end{eqnarray}
	\end{subequations} having introduced the \text{We}ber number $\text{\text{We}}=\sigma/(\rho g [h]^2)$ for the FF problem, as in \citet{falling_liquid_films}, and the capillary number $\text{Ca}=\mu U_p/\sigma$ for the MS problem, as in \citet{mendez_jfm}. Moreover, it is convenient to introduce a reduced Reynolds number, defined as $\delta=\varepsilon \text{Re}$, and a dimensionless number that combines $\text{Re}$ and $\text{\text{We}}$ (or $\text{Ca}$) and depends only on the liquid properties. This is the Kaptiza number $\text{Ka}=\sigma/(\rho \nu^{4/3} g^{1/3})$, which weighs the importance of surface tension over viscosity (see also \citet{Mendez2017}). The remaining set of reference quantities is finally $[v]=\varepsilon [u]$ and $[t]=[x]/[u]$, taking the appropriate references for the FF and the MS problems.

	\section{Integral Boundary Layer Models}\label{sec3}
	
	In both the FF and the MS problems, the integral boundary layer (IBL) models can be derived from the Navier-Stokes equations, scaled according to the reference quantities in Section \ref{sec2} and retaining only terms up to $\mathcal{O}(\varepsilon)$ (see Appendix \ref{app:NS} for details). This results in the boundary layer equations:
	\begin{subequations}
		\label{eq:dimensionless-model}
		\begin{eqnarray}
			\partial_{\hat{x}} \hat{u} + \partial_{\hat{y}} \hat{v} + \partial_{\hat{z}} \hat{w} &=& 0, \label{eq:dimensionless-cont} \\
			\delta \big( \partial_{\hat{t}} \hat{u} + \hat{u} \partial_{\hat{x}} \hat{u} + \hat{v} \partial_{\hat{y}} \hat{u} + \hat{w} \partial_{\hat{z}} \hat{u} \big) &=& -\partial_{\hat{x}} \hat{p_x} \nonumber \\ && + \partial^2_{\hat{y}\hat{y}} \hat{u} + 1, \label{eq:dimensionless-xmom}\\
			\partial_{\hat{y}} \hat{p_y} &=& 0, \label{eq:dimensionless-ymom} \\
			\delta \big( \partial_{\hat{t}} \hat{w} + \hat{u} \partial_{\hat{x}} \hat{w} + \hat{v} \partial_{\hat{y}} \hat{w} + \hat{w} \partial_{\hat{z}} \hat{w} \big) &=& -\partial_{\hat{z}} \hat{p_z} 
			+ \partial^2_{\hat{y}\hat{y}} \hat{w}, \label{eq:dimensionless-zmom}
		\end{eqnarray}
	\end{subequations}
	where \eqref{eq:dimensionless-cont} is the continuity equation, and \eqref{eq:dimensionless-xmom}, \eqref{eq:dimensionless-ymom}, \eqref{eq:dimensionless-zmom} are the momentum equations along $\hat{x}$, $\hat{y}$, and $\hat{z}$ respectively. 
	The hat indicates dimensionless quantities.
	The dimensionless kinematic boundary conditions at the wall and at the interface are
	\begin{subequations}
		\label{eq:dimensionless-bc}
		\begin{eqnarray}
			\vec{\hat{v}} \big|_{\hat{y}=0} &=& (\hat{u},\hat{v},\hat{w}) \big|_{\hat{y}=0} = (\alpha, 0, 0), 
			\label{eq:kin-bc} \\
			\hat{v} \big|_{\hat{y}=\hat{h}} &=& \partial_{\hat{t}} \hat{h} + \hat{u} \big|_{\hat{y}=\hat{h}} \partial_{\hat{x}} \hat{h} + \hat{w} \big|_{\hat{y}=\hat{h}} \partial_{\hat{z}} \hat{h},
		\end{eqnarray}
	\end{subequations} where $\alpha=0$ for the FF problem and $\alpha=-1$ 
	for the MS problem. This parameter is introduced to link the derivation of the two models, but it is worth stressing that these problems have different scaling laws, as described in Section \ref{sec2}. At $\mathcal{O}(\varepsilon)$, the dynamic boundary conditions formulating the force balance at the free surface is:
	\begin{subequations}
		\label{eq:dimensionless-bc-dynamic}
		\begin{eqnarray}
			\hat{p} \big|_{\hat{y}=\hat{h}} &=& \hat{p}_g - (\partial_{\hat{x}\hat{x}} \hat{h} + \partial_{\hat{z}\hat{z}} \hat{h} ), \\
			\partial_{\hat{y}} \hat{u} \big|_{\hat{y}=\hat{h}} &=& \hat{\tau}_{g,x}, \\
			\partial_{\hat{y}} \hat{w} \big|_{\hat{y}=\hat{h}} &=& \hat{\tau}_{g,z},
		\end{eqnarray}
	\end{subequations} where $\hat{p}_g$, $\hat{\tau}_{g,x}$ and $\hat{\tau}_{g,z}$ are the gas pressure and the shear stress components along $x$ and $z$ respectively, imposed by an external air flow.
	
	To derive the integral model, we integrate \eqref{eq:dimensionless-model} along $y$ assuming a self-similar parabolic velocity profile for both the streamwise $\hat{u}$ and the spanwise $\hat{w}$ velocity components, as in \citet{demekhin_shkadov}. Using the local flow rate definitions, the substrate motion and the interface shear stress, the profiles for the MS case read: 
	\begin{subequations}
		\begin{eqnarray}
			\hat{u} (\hat{h}, \hat{q}_x, \hat{q}_z) &=& \frac{3}{4\hat{h}^3}\big( \hat{\tau}_{g,x} \hat{h}^2 - 2\hat{h} - 2\hat{q}_x \big) \hat{y}^2 \nonumber \\ && + \frac{6\hat{h} + 6\hat{q}_x-\hat{\tau}_{g,x}\hat{h}^2}{2\hat{h}^2} \hat{y} - 1, \\
			\hat{w} (\hat{h}, \hat{q}_x, \hat{q}_z) &=& \frac{3}{4\hat{h}^3}\big( \hat{\tau}_{g,z} \hat{h}^2 - 2\hat{q}_z \big) \hat{y}^2 \nonumber \\ && + \frac{6\hat{q}_z-\hat{\tau}_{g,z}\hat{h}^2}{2\hat{h}^2} \hat{y}.
		\end{eqnarray}
	\end{subequations}
	The main hypothesis under this assumption is that the balance of viscosity and gravity is not significantly altered by inertia and surface tension. 
	
	The integration results in a system of nonlinear partial differential equations for the liquid film height $\hat{h}$, the streamwise $\hat{q}_x$ and spanwise $\hat{q}_z$ flow rates. In conservative form, this reads:
	\begin{eqnarray}
		\label{eq:model_gen}
		\partial_{\hat{t}} \vec{U} + \nabla \cdot \mathbf{F} = \vec{S},
	\end{eqnarray}
	with the state vector $\vec{U}$ consisting of the liquid film height and the volumetric flow rates, $ \vec{U}  = (\hat{h}, \hat{q}_x, \hat{q}_z)^\text{T}$.
	The source vector is denoted by $\vec{S} = (S_1, S_2, S_3)^\text{T}$, and reads
	\begin{eqnarray}
		\label{eq:sources}
		\vec{S} &=&  
		\begin{pmatrix}
			0 \\
			\frac{1}{\delta} \Big[ \hat{h} \Big( - \partial_{\hat{x}} \hat{p}_x + \partial_{\hat{x}\hat{x}\hat{x}} \hat{h} + \partial_{\hat{x}\hat{z}\hat{z}} \hat{h} + 1 \Big) + \Delta \hat{\tau}_{x} \Big] \\
			\frac{1}{\delta} \Big[ \hat{h} \Big( - \partial_{\hat{z}} \hat{p}_z + \partial_{\hat{z}\hat{z}\hat{z}} \hat{h} + \partial_{\hat{z}\hat{x}\hat{x}} \hat{h} \Big) + \Delta \hat{\tau}_{z} \Big]
		\end{pmatrix}
	\end{eqnarray}
	where the terms with third derivatives of $\hat{h}$ correspond to the capillary pressure gradients, and
	the terms $\Delta \hat{\tau}_{x} = \hat{\tau}_{g,x} + \hat{\tau}_{w,x}$ and $\Delta \hat{\tau}_{z} = \hat{\tau}_{g,z} + \hat{\tau}_{w,z}$ result from the integration of the viscous terms in \eqref{eq:dimensionless-xmom} and \eqref{eq:dimensionless-zmom}. These represent the difference in shear stress between the interface (terms $\hat{\tau}_{g,x}$ and $\hat{\tau}_{g,z}$) and the wall (terms $\hat{\tau}_{w,x}$ and $\hat{\tau}_{w,z}$).
	The shear stress at the wall, using the self-similar assumption for the velocity profiles, reads:
	\begin{subequations}
		\label{Shear_WALL}
		\begin{eqnarray}
			\hat{\tau}_{w,x} &=& \frac{1}{2} \hat{\tau}_{g,x}  -\frac{3\hat{q}_x}{\hat{h}^2} + \alpha \frac{3}{\hat{h}}, \\
			\hat{\tau}_{w,z} &=& \frac{1}{2} \hat{\tau}_{g,z} -\frac{3\hat{q}_z}{\hat{h}^2}.
		\end{eqnarray}
	\end{subequations}
	
	The flux matrix $\mathbf{F}$ in \eqref{eq:model_gen} is 
	\begin{eqnarray}
		\mathbf{F} &=& 
		\begin{pmatrix} 
			F_{11} & F_{12} & F_{13} \\
			F_{21} & F_{22} & F_{23} 
		\end{pmatrix} = 
		\begin{pmatrix} 
			\hat{q}_x & \int_0^{\hat{h}} \hat{u}^2 d\hat{y} & \int_0^{\hat{h}} \hat{u} \hat{w} d\hat{y} \\
			\hat{q}_z & \int_0^{\hat{h}} \hat{u} \hat{w} d\hat{y} & \int_0^{\hat{h}} \hat{w}^2 d\hat{y}
		\end{pmatrix}
	\end{eqnarray}
	and has the following components:
	\begin{subequations}
		\allowdisplaybreaks
		\label{eq:fluxes_terms}
		\begin{eqnarray}
			F_{11} &=& \int_0^{\hat{h}} \hat{u} d \hat{y} \eqqcolon \hat{q}_x, 
			\\[0.3em]
			F_{21} &=& \int_0^{\hat{h}} \hat{w} d \hat{y} \eqqcolon \hat{q}_z, 
			\\[0.3em]
			F_{12} &=& \frac{1}{120 \hat{h}} \Big(144 \hat{q}_x^2 + 6 \hat{\tau}_{g,x} \hat{h}^2 \hat{q}_x + \hat{\tau}_{g,x} \hat{h}^4 \nonumber \\&& 
			-\alpha \big(48 \hat{h} \hat{q}_x + 6 \hat{\tau}_{g,x} \hat{h}^3 + 24 \hat{h}^2\big) \Big), 
			\\[0.3em]
			F_{22} &=& \frac{1}{120 \hat{h}} \Big( 144 \hat{q}_x \hat{q}_z + 3 \hat{\tau}_{g,x} \hat{h}^2 \hat{q}_z + 3 \hat{\tau}_{g,z} \hat{h}^2 \hat{q}_x  \nonumber \\&& 
			+ \hat{\tau}_{g,x} \hat{\tau}_{g,z} \hat{h}^4
			-\alpha \big( 24 \hat{h} \hat{q}_z + 3 \hat{\tau}_{g,z} \hat{h}^3\big) \Big), 
			\\[0.3em]
			F_{13} &=& F_{22}, 
			\\[0.3em]
			F_{23} &=& \frac{144 \hat{q}_z^2 + 6 \hat{\tau}_{g,z} \hat{h}^2 \hat{q}_z + \hat{\tau}_{g,z}^2 \hat{h}^4}{120 \hat{h}}. 
		\end{eqnarray}
	\end{subequations}
	
	This model recovers the 3-D model for falling liquid films by \citet{demekhin_2007} if $\alpha=0$, $\partial_{\hat{x}} \hat{p}_x=\partial_{\hat{x}} \hat{p}_z=0$, $\hat{\tau}_{g,x}=\hat{\tau}_{g,z}=0$, and its two-dimensional version by \citet{Shkadov1970} if also $\hat{q}_z=0$ and $\partial_{z}\rightarrow 0$. Moreover, the model recovers the 2-D liquid film model in jet wiping by \citet{mendez_jfm} if $\alpha=-1$, $\hat{q}_z=0$, $\hat{\tau}_{g,z}=0$ and $\partial_{z} \rightarrow 0$. If $\alpha=-1$ is introduced in system \eqref{eq:model_gen}, we obtain the first 3-D formulation of an integral boundary layer model for the jet wiping process.

	\section{Linear Stability Analysis in 2-D}\label{sec4}
	\label{sec:Lin stab}
	We consider the 2-D linear stability analysis of the IBL model for FF and MS conditions, i.e. setting $\hat{q}_z=0$ in the system \eqref{eq:model_gen} 
	and assuming $\partial_{\hat{z}} \rightarrow 0$, no shear stress at the interface ($\hat{\tau}_{g,x}=\hat{\tau}_{g,z}=0$) and no pressure gradient ($\partial_{\hat{x}} \hat{p}_g=0$). We then introduce
	\begin{equation}
		\hat{h} = \hat{h}_0 + \tilde{h}, \quad \hat{q}_{x} = \hat{q}_0 + \tilde{q},
	\end{equation}
	in the governing Eq. \eqref{eq:model_gen},
	with $\hat{h}_0,\hat{q}_0$ denoting the thickness and flow rates at an equilibrium solution and  $\tilde{h}<<\hat{h}_0$, $\tilde{q}<<\hat{q}_0$ some small perturbations. Linearizing around $\hat{h}_0, \hat{q}_0$ yields the perturbation equations:
	\begin{subequations}
		\label{eq:model_linear}
		\begin{eqnarray}
			\partial_{\hat{t}} \tilde{h} + \partial_{x} \tilde{q} &=& 0,
			\label{eq:model_linear_a} 
			\\
			\delta\Big(\partial_{\hat{t}} \tilde{q} + \partial_{\hat{x}} \tilde{F}\Big) &=& (\hat{h}_0 + \tilde{h}) + (\hat{h}_0 + \tilde{h})\partial_{\hat{x}\hat{x}\hat{x}}\tilde{h} + \Delta\tilde{\tau}, \label{eq:model_linear_b}
		\end{eqnarray}
	\end{subequations}
	with
	\begin{subequations}
		\begin{eqnarray}
			\tilde{F} &=& \frac{6(\hat{q}_0^2 + 2\hat{q}_0\tilde{q})}{5(\hat{h}_0 + \tilde{h})} 
			- \alpha\Biggl[ \frac{2}{5} (\hat{q}_0+\tilde{q})+\frac{1}{5}(\hat{h}_0+\tilde{h})\Biggr]
			\\
			\Delta\tilde{\tau} &=& - \frac{3(\hat{q}_0 + \tilde{q})}{\hat{h}_0^2 + 2\hat{h}_0\tilde{h}} + \frac{3\alpha}{\hat{h}_0 + \tilde{h}}\,.
		\end{eqnarray}
	\end{subequations}

	We now consider a perturbation in the form of a normal mode, hence:
	\begin{equation}
		\tilde{h} = h_{\varepsilon}\exp\bigl[i(\hat{k}\hat{x}-\hat{\omega} \hat{t})\bigr], \; \tilde{q} = q_{\varepsilon}\exp\bigl[i(\hat{k}\hat{x}-\hat{\omega} \hat{t})\bigr],
		\label{normal_modes}
	\end{equation}
	where $\hat{k}$ is the dimensionless wave number and $\omega=\hat{\omega}_r + i\hat{\omega}_i$ is the complex dimensionless angular frequency. 
	Substituting Eq. \eqref{normal_modes} into Eq. \eqref{eq:model_linear_b}, noticing that the base state leads to $\hat{q}_0=\hat{h}^3_0/3+\alpha \hat{h}_0$ and separating real and imaginary parts gives an algebraic system of nonlinear equations:
	\begin{subequations}
		\label{General_STABILITY}
		\begin{eqnarray}
			&& \delta\Big[2\hat{\omega}_r\hat{\omega}_i\hat{h}_0^2 - \frac{12}{5}\hat{q}_0\hat{h}_0\hat{k}\hat{\omega}_i +\alpha\frac{2}{5}\hat{h}_0^2\hat{k}\hat{\omega}_i\Big] \nonumber \\ && \qquad - 3k\hat{h}_0^2 + 3\hat{\omega}_r -3\alpha\hat{k}=0,
			\label{eq:17a}\\
			&& \delta\Big[(\hat{\omega}_r^2-\hat{\omega}_i^2)\hat{h}_0^2 - \frac{12}{5}\hat{q}_0\hat{h}_0\hat{k}\hat{\omega}_r + \frac{6}{5}k^2\hat{q}_0^2\Big] \nonumber \\ && \qquad + \delta\alpha\Big(\frac{2}{5}\hat{h}_0^2\hat{k}\hat{\omega}_r + \frac{\hat{k}^2\hat{h}_0^2}{5}\Big) \nonumber \\ && \qquad - \hat{k}^4\hat{h}_0^3 - 3 \hat{\omega}_i = 0. \label{eq:17b}
		\end{eqnarray}
	\end{subequations}
	
	Since $\hat{q}_0$ and $\hat{h}_0$ are linked by $\hat{q}_0=\hat{h}^3_0/3+\alpha \hat{h}_0$, for a given $\hat{h}_0$ and a pair of $(\hat{k},\delta)$, these equations can be solved for $\hat{\omega}_i,\hat{\omega}_r,\hat{k}$. Moreover, by setting $\hat{\omega}_i=0$, we find the dispersion relation and the neutral curves (i.e. the loci of conditions in which disturbances neither grow nor decay).
	
	In the FF problem, one has $\alpha=0$ and $\hat{h}_0 = 1, \hat{q}_0 = 1/3$. The neutral curve is 
	\begin{subequations}
		\begin{eqnarray}
			\hat{\omega}_r = \hat{k} && \quad (\text{Dispersion Relation}) \\
			\hat{k} = \sqrt{\frac{\delta}{3}} && \quad (\text{Neutral Curve})\,.
		\end{eqnarray}
	\end{subequations}
	
	In the MS problem, one has $\alpha=-1$ and any $\hat{h}_0\in[0,\sqrt{3}]$ is a possible steady state solution (see \citet{mendez_jfm}). The neutral curve is 
	\begin{subequations}
		\begin{eqnarray}
			\hat{\omega}_r = \Big(\hat{h}_0^2 - 1\Big)\hat{k} &&  \quad (\text{Dispersion Relation}) \label{eq:19a_disp_rel} \\
			\hat{k} =  \sqrt{\frac{h_0^3\delta}{3}} && \quad (\text{Neutral Curve})\,. \label{eq:19b_neutral_curve}
		\end{eqnarray}
	\end{subequations}
	Interestingly, at $\hat{h}_0=1$ one has $\hat{\omega}_r=0$ for all wave numbers.

	\section{Numerical Methods} \label{sec5}
	
	We developed an in-house finite volume solver in Python. This is a 3-D extension of the 2-D solver in \citet{mendez_jfm}. 
	More specifically, we blend the two-steps Lax-Wendroff and the two-steps Lax-Friedrichs schemes by \citet{Shampine2005}. The blending is carried out using flux limiters. These allow for switching between second-order (Lax-Wendroff) and first-order (Lax-Friedrichs) accuracy depending on the steepness of the solution. 
	More details on the solver and the discretization schemes can be found in Appendix \ref{app:num}.

	\subsection{Validation of the solver for the FF problem}
	\label{sec5p1}
	
	We begin by considering a test case of the FF problem. 
	The test case is from \citet{Doro2013}, who presented a numerical investigation of falling liquid films using the Volume of Fluid (VOF) solver in OpenFOAM. In this test case, the liquid is an aqueous solution of dimethylsulfoxide (DMSO)  
	at Reynolds number $\text{Re} = 15$ and Kapitza number $\text{Ka} = 509$. The liquid properties are $\rho = 1098.3$ \si{\kg \per \metre \cubed}, $\nu = 2.85 \times 10^{-6}$ \si{\metre\squared\per\second} and $\sigma = 0.0484$ \si{\newton \per \metre}.
	The domain length in our simulation covers $0.153$ \si {\metre}. In dimensionless units, this yields $L_x = 140$, with cell size set to $dx = dz = 0.1$. The time step is set to $dt = 0.01$.
	At the domain's inlet, perturbations to the flow rate are introduced at a frequency $f = 16$ \si{\hertz}. Thus, the inlet conditions are:
	\begin{eqnarray}
		\label{eq:pert_fixed_plate}
		\hat{q}_x &=& \frac{1}{3} \hat{q}_A \sin(2 \pi \hat{f}\hat{t} ) + \frac{1}{3}, \\ \nonumber
		\hat{h} &=& (3 \hat{q}_x)^{1/3},
	\end{eqnarray}
	where $\hat{q}_A = 0.05$ is the perturbation amplitude, $\hat{f} = 12[t]=0.048$ is the dimensionless frequency, and $\hat{t} = n dt$ is the dimensionless time stepping of the simulation.
	
\begin{figure}[htbp]
		\centering
		\includegraphics[width=1\linewidth]
		{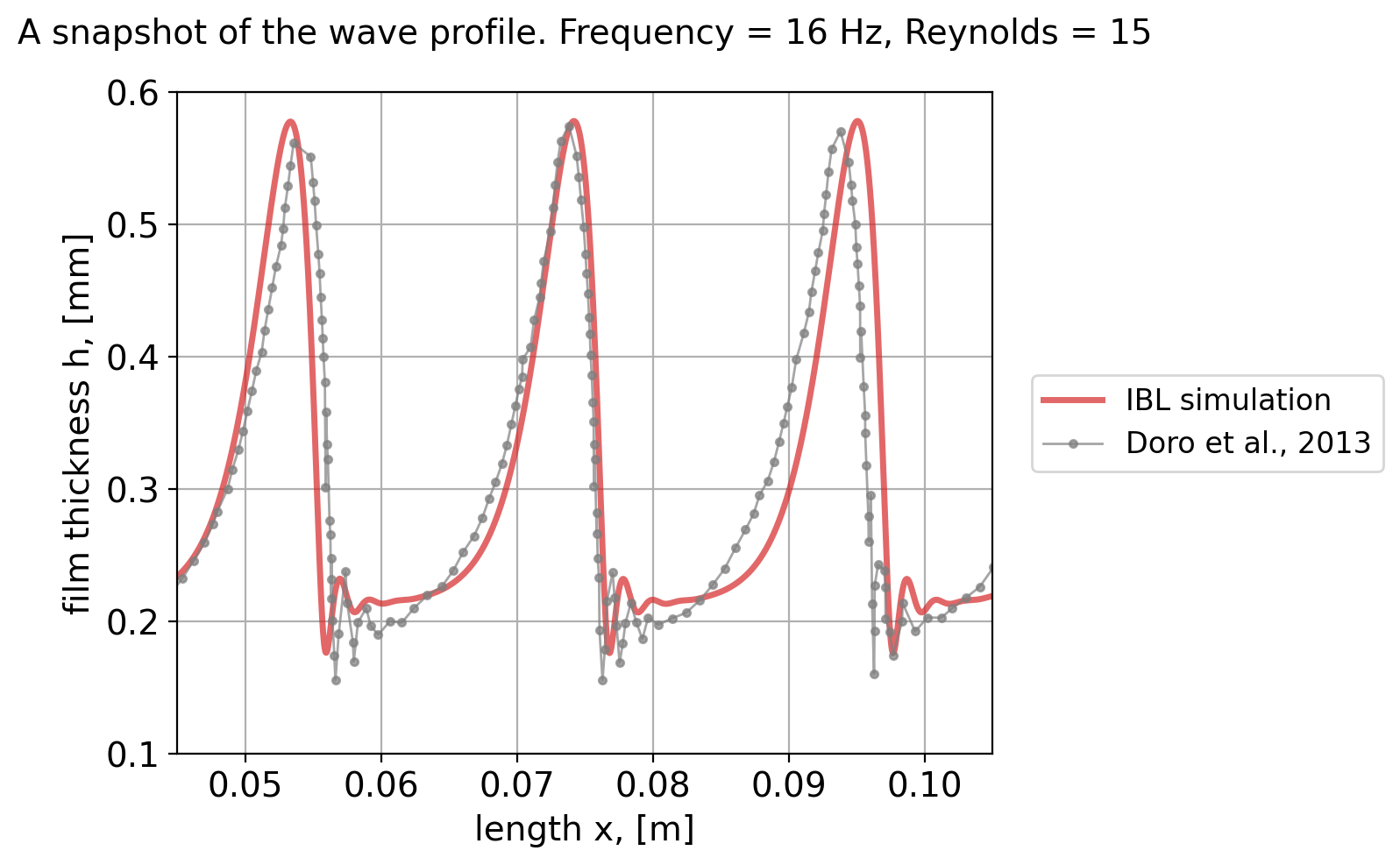}
		\caption{Simulation with the IBL solver of a DMSO falling film to validate it with results from \citet{Doro2013} for the FF problem.}
		\label{fig:doro}
	\end{figure}

	The disturbances grow over the domain until they produce the classic wave train observed in forced flows. Sufficiently far from the inlet, the waves are developed and their shape and phase is nearly invariant to the streamwise location, as discussed in \citet{Doro2013}. For this region, the comparison between the IBL simulations and the VOF simulations (from Fig. 5 in \citet{Doro2013}) is shown in Fig. \ref{fig:doro}. 
	
	We observe a good agreement between the two simulations, although it is over-simplifying to assume self-similarity and $\mathcal{O}(\varepsilon)$ accuracy in the modeling of a problem that has $\varepsilon=0.29$.
	Nevertheless, considering that the computational cost of a (1D) IBL simulation is several orders of magnitude lower than the computational cost of a (2-D) VOF simulation, the result is particularly encouraging.
	
	\subsection{Validation of the solver for the MS problem}
	\label{sec5p2}
	
	For the MS problem, we consider the same validation test case as in \citet{mendez_jfm} to validate our solver. This test case consists of a 2-D wave train over a moving substrate and was also simulated using high-fidelity VOF simulations in OpenFOAM. The liquid is water with a Reynolds number $Re = 319$ and the substrate moves at $U_p=1$ \si{\metre\per\second}.
	
	The computational domain is rectangular, with a dimensionless length $L_x = 8400 h_0$ in the streamwise direction, and $L_z = 7.8 h_0$ in the spanwise direction, small enough to keep 2-D waves, i.e. stable with respect to spanwise perturbations. A perturbation with a dimensionless frequency $\hat{f} = 0.05$ is introduced at the inlet's streamwise flow rate:
	\begin{eqnarray}
		\label{eq:pert}
		\hat{q}_x = \big[\frac{1}{3} \hat{h}_0^3 - \hat{h}_0\big] \big[1+\hat{q}_A \sin(2 \pi \hat{f}\hat{t} )\big],
	\end{eqnarray}
	where $\hat{q}_A$ is the perturbation amplitude.
	
\begin{figure}[htbp]
		\subfloat[Comparison of the solutions obtained with the Lax-Friedrichs scheme for different cell sizes.]{\label{fig:OF_mesh_sensitivity}\includegraphics[width=\linewidth]{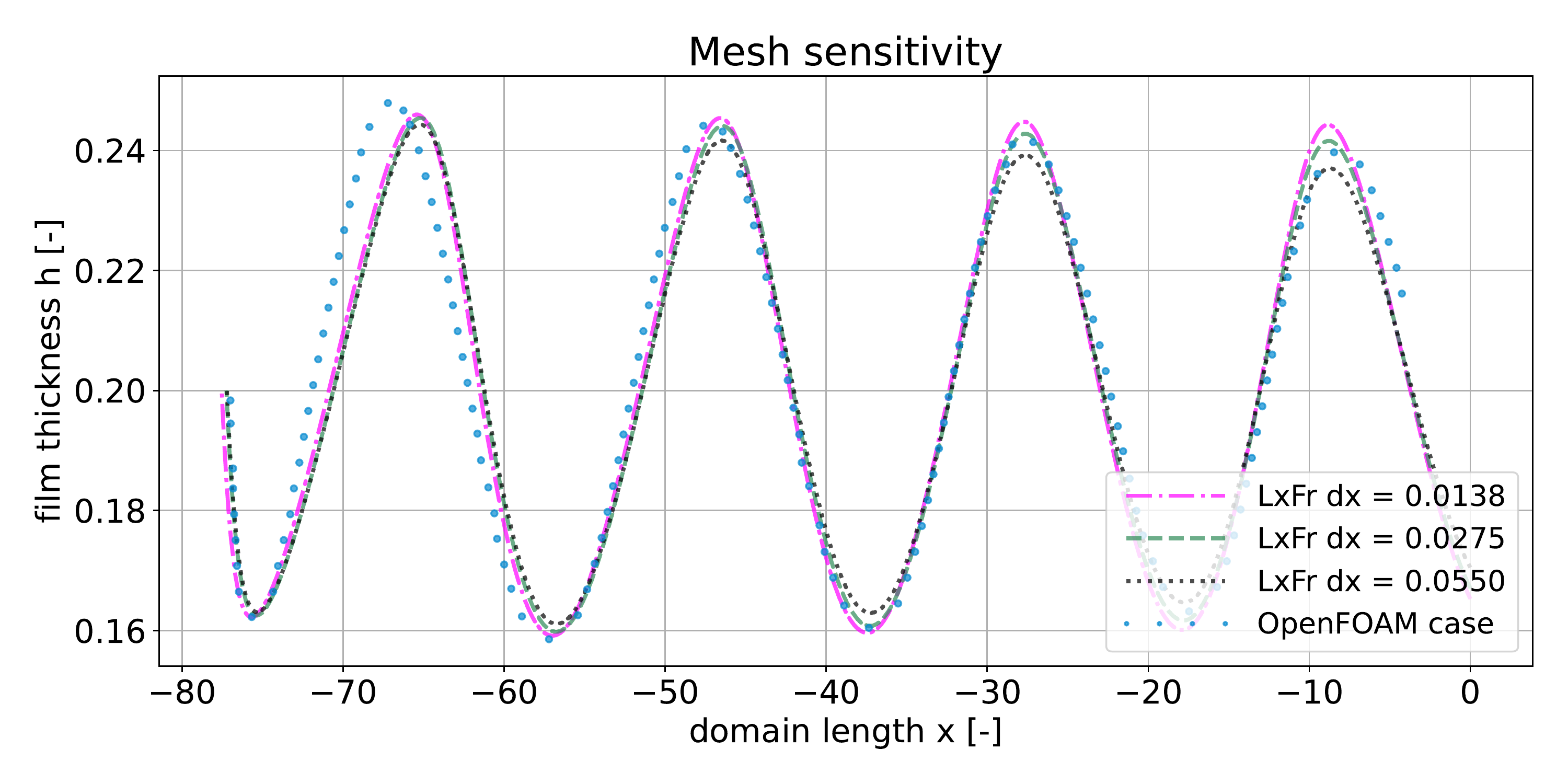}}
		
		\subfloat[Validation of the current solver (with its solution in a dash-dotted line) against the OpenFOAM case, which is also used in \citet{mendez_jfm}.]{\label{fig:OF_validation}\includegraphics[width=1\linewidth]{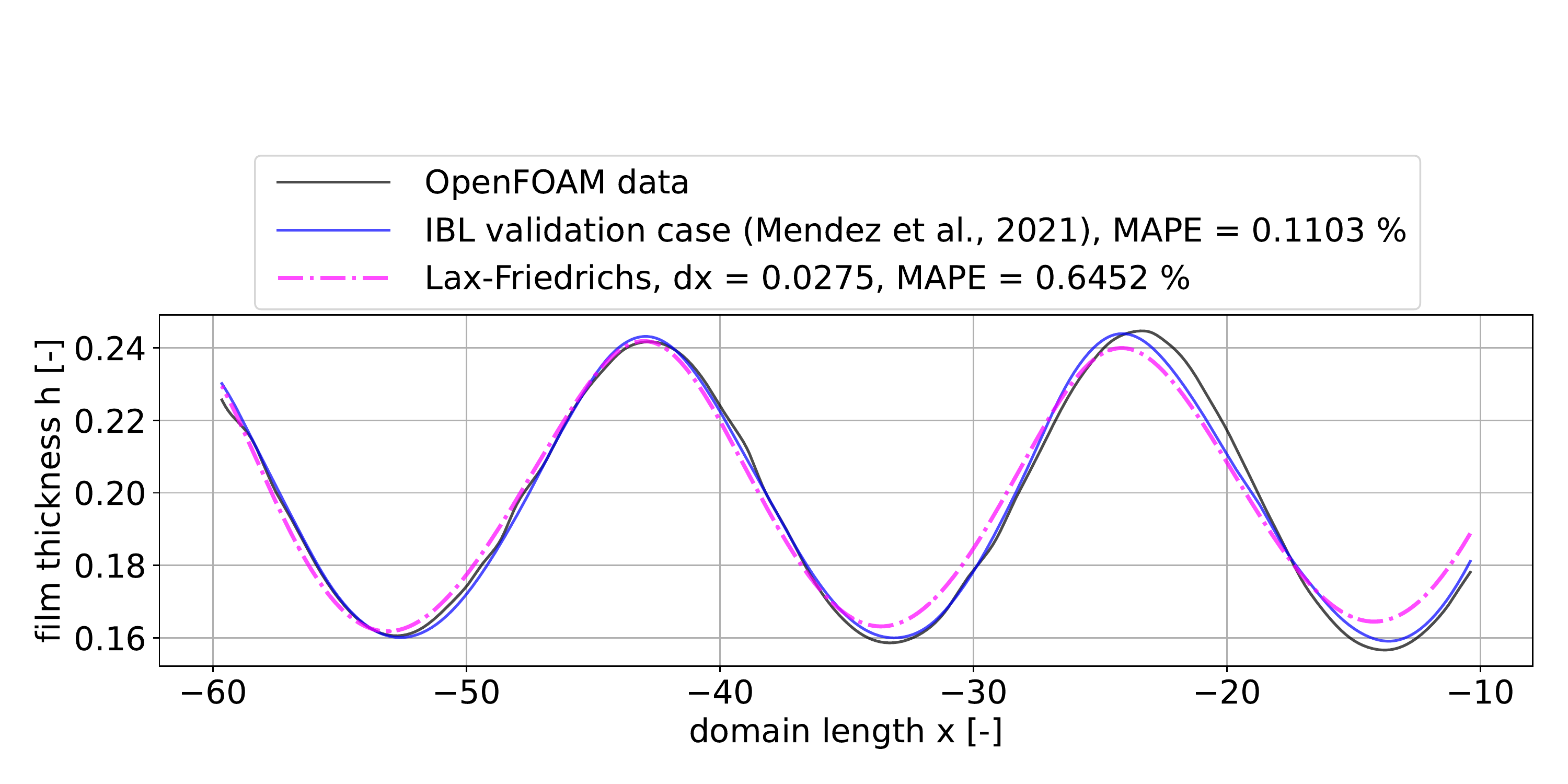}}
		
		\caption{Validation of the solver for the MS case. The inlet is at $\hat{x}=80$ and the substrate moves against the direction of gravity, which is $\hat{x} \to \infty$. MAPE is the mean absolute percentage error between the solution with Lax-Friedrichs (dot-dashed line) and the OpenFOAM data.}
		\label{fig:validation}
	\end{figure}

	This test case was also used to perform a mesh sensitivity analysis of our solver. This analysis was based on solutions we obtained with the Lax-Friedrichs scheme for three different cell sizes $d\hat{x}$, namely $d\hat{x}=0.0138$, $d\hat{x}=0.0275$ and $d\hat{x}=0.0550$. The results for the thickness evolution in these three cases are shown in Fig. \ref{fig:OF_mesh_sensitivity} and compared to the results obtained by the VOF simulations in OpenFOAM (in which $d\hat{x}=0.0275$). Fig. \ref{fig:OF_validation} further compares the obtained solution for $d\hat{x}=0.0275$ with the OpenFOAM validation case from \citet{mendez_jfm}.
	While numerical dissipation is visible and it is larger for coarser meshes as expected, its impact can be considered minor within the investigated domain.  
	
	
	\vspace{-3mm}
	\subsection{Investigated test cases}
	\label{sec5p3}
	
	We analyzed the propagation of nonlinear waves in the MS problem keeping the same configuration used for the solver validation in Section \ref{sec5p2}, hence introducing perturbations at the inlet flow rate as in Eq. \eqref{eq:pert}. We consider three dimensionless thicknesses $\hat{h}_0=0.1$, $0.2$, $0.3$, two reduced Reynolds numbers $\delta_1=76$ and $\delta_2=153$, and fourteen dimensionless frequencies in the range $\hat{f}=[0.005,0.2]$. This leads to 84 simulations. The simulations were carried out without surface tension to focus on the impact of nonlinearities in the interface instability. Without surface tension linear stability theory predicts that all configurations are unstable, as shown in Section \ref{sec6p1}. However, this was found not to be the case if nonlinearities are considered.
	
	
	The dimensionless conditions are representative for hot dip galvanizing lines as well as the laboratory model ESSOR at the von Karman institute (see \citet{Buchlin,Gosset2019,mendez_jfm}). The similarity between water and molten zinc in the Skhadov-like scaling used in this work was discussed in \citet{mendez_jfm}. For a plate moving at $U_p=1$ \si{\m\per\s}, taking $\rho = 1000$ \si{\kg \per \metre \cubed}, $\nu = 1$ \si{\mm \squared \per \s}, $\sigma=0.074$ \si{\N \per \m} for water, and $\rho = 6500$ \si{\kg \per \metre \cubed}, $\nu = 0.46$ \si{\mm \squared \per \s}, $\sigma=0.78$ \si{\N \per \m} for molten zinc leads to $\delta\approx 76$ for both fluids.
	
	In all test cases, waves propagate in the direction of the strip motion ($\hat{x}\rightarrow -\infty$, cf. Fig. \ref{fig:jw3d}). Therefore, the domain was set as $\hat{x}\in[0,40]$, with initial disturbance placed at $\hat{x}=40$. 
	Both the size of the domain and the mesh size vary from test case to test case. Specifically, considering that waves propagate at about $\hat{u}_{\text{w}} \approx 1$ and their wavelength is of the order of
	$\hat{\lambda} \approx 1/\hat{f}$, the domain length in the streamwise direction is taken as $L_x = 8 \lambda$ while the width is taken as
	$L_z \approx L_x / 10$. The grid spacing is taken as $d x = \lambda / 363$ and $d z = 1/100$, since $363$ points per wavelength proved to give a good compromise between accuracy and computational cost. The time step is taken such
	that the Courant–Friedrichs–Lewy (CFL) number is 0.3, using $|u_w|=1$ as an estimate of the wave velocity. This yields a numerical viscosity
	$\mu_n = d \hat{x}^2 / d \hat{t} \propto d \hat{x}$.
	
	Finally, we analyzed the spanwise propagation of three-dimensional disturbances by considering an inlet flow rate $\hat{q}_x$ consisting of an harmonic term modulated by a Gaussian function $G$ along $\hat{z}$ 
	\begin{equation}
		\label{eq:pert_qx}
		\hat{q}_x = \big[\frac{1}{3} \hat{h}_0^3 - \hat{h}_0\big] \Big[1+\hat{q}_A \sin(2 \pi \hat{f}\hat{t}) \sin \Big( \frac{2 \pi}{\lambda_z} \hat{z} \Big) \Big] G(\hat{z}),
	\end{equation} where $\hat{\lambda}_z=1$ and the Gaussian modulation with a standard deviation $\sigma = 0.4$ was taken as 
	\begin{equation}
		G(\hat{z}) = \frac{1}{\sigma \sqrt{2 \pi}} \exp \Big( \frac{-(z-z_{\text{mean}})^2}{2 \sigma^2} \Big).
	\end{equation}
	
	For the inlet flow rate along $z$, we consider $\hat{q}_z = 0$. This yields $\hat{q}_z=0$ everywhere and at all times.
	
	\section{Results and Discussions} 
	\label{sec6}
	
	\subsection{Linear stability}
	\label{sec6p1}
	
	We first consider the result from a linear stability analysis following equations \eqref{General_STABILITY} for both FF and MS. Given a computational grid of 1000 dimensionless wave numbers $\hat{k}$ and 1000 reduced Reynolds numbers $\delta$, Eq. \eqref{General_STABILITY} were solved for $\hat{\omega}_r$ and $\hat{\omega}_i$ for each pair $(\hat{k},\delta)$. Because multiple solutions exist, we only focus on the largest $\hat{\omega}_i$ (regardless of the sign) and plot a contour of $\log(\hat{\omega}_i)$ on the grid $(\hat{k},\delta)$. 
	
	\begin{figure}[htbp]
		\centering
		\includegraphics[width=\linewidth]{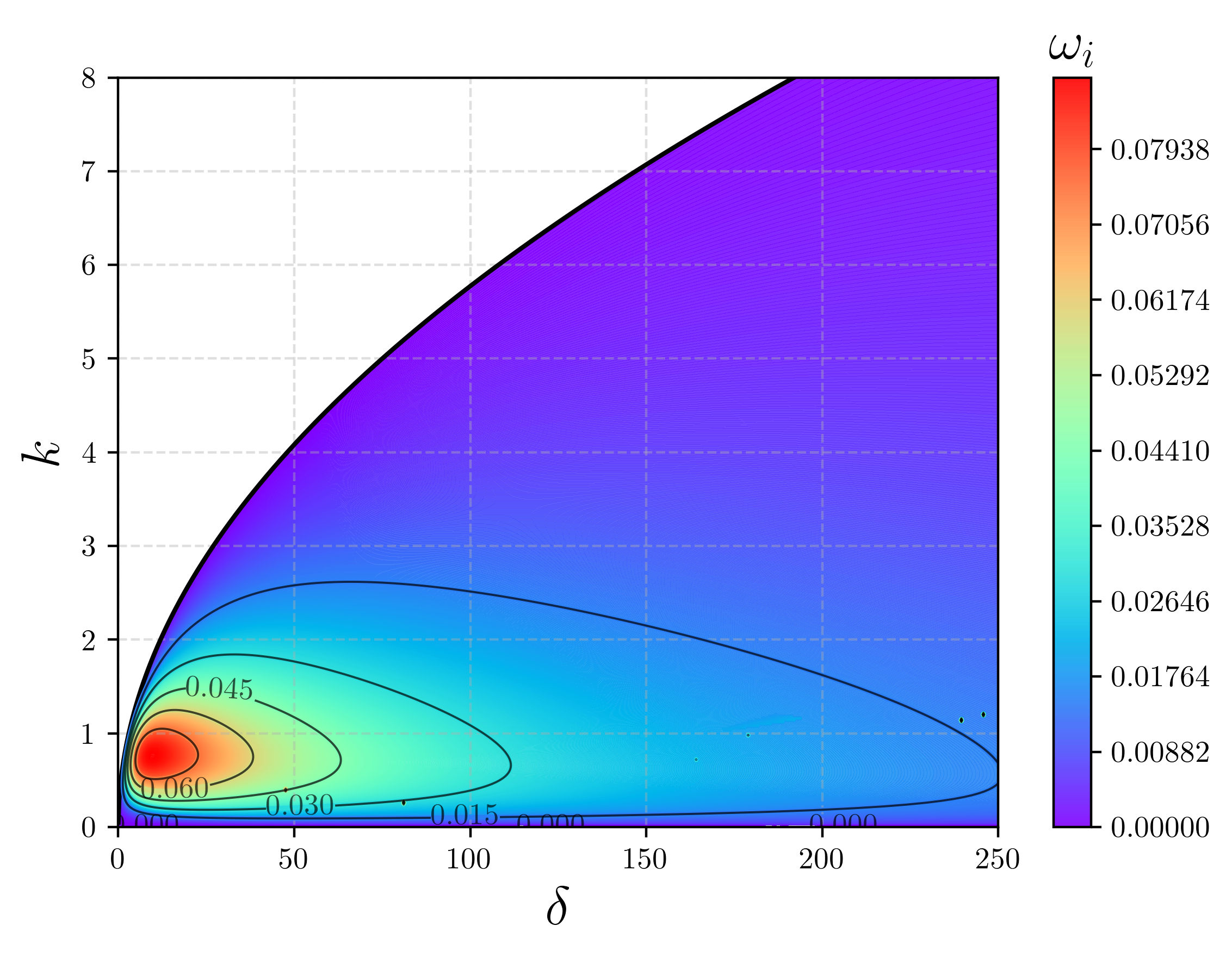}
		\caption{Color map of $\log(\hat{\omega}_i)$ with the natural stability curve (black thick line) for the FF case.}
		\label{fig:omega:FF}
	\end{figure}

	For the FF problem one has $\alpha=0$, $\hat{h}_0=1$ and $\hat{q}_0=1/3$. The associated amplification factors are shown in Fig. \ref{fig:omega:FF}. The region in white in the $(k,\delta)$-plane is the region where $\hat{\omega}_i<0$, hence where the film is stable according to the linear stability analysis. The line separating the stable and unstable regions is the neutral stability curve. For $k<k_c$, the amplification factors show a non-monotonic dependence on $\delta$: while the region of maximum amplification occurs in $\delta\approx 16$, increasing the Reynolds number leads to reduction of $\omega_i$ for all wavelengths. This trend is due to the scaling of the problem, since $[\omega]=1/[t]\propto U_p^{5/6}\propto \delta^{5/11}$. For later reference, at $\delta=76$ and $\delta=153$ the critical wave-numbers are $k_c\approx5$ and $k_c\approx 7.1$ respectively.
	
	\begin{figure}[htbp]
		\centering    
		\subfloat[Case for $\hat{h}_0=0.1$.]{\includegraphics[width=0.98\linewidth]{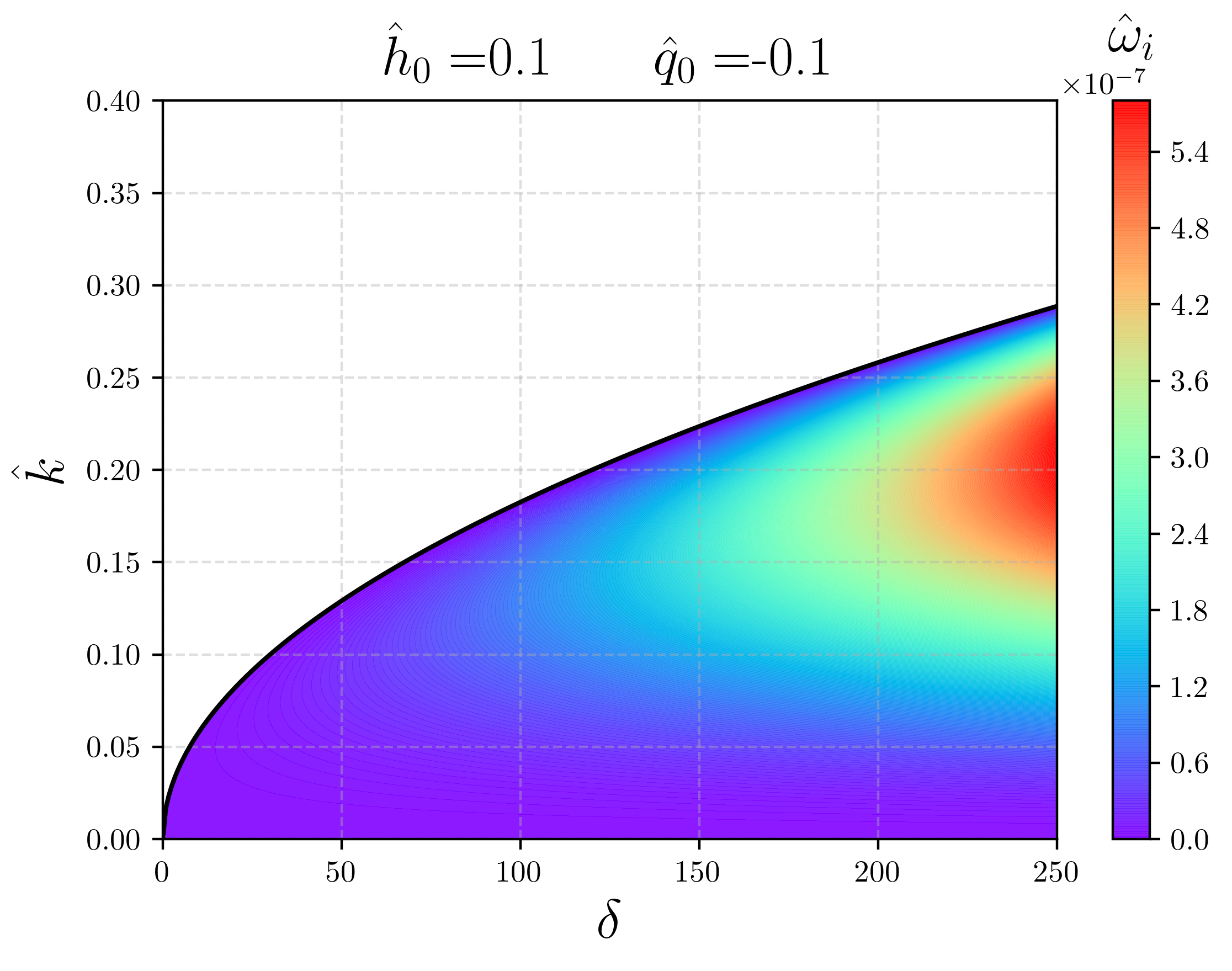}}
		
		\subfloat[Case for $\hat{h}_0=0.2$.]{\includegraphics[width=0.98\linewidth]{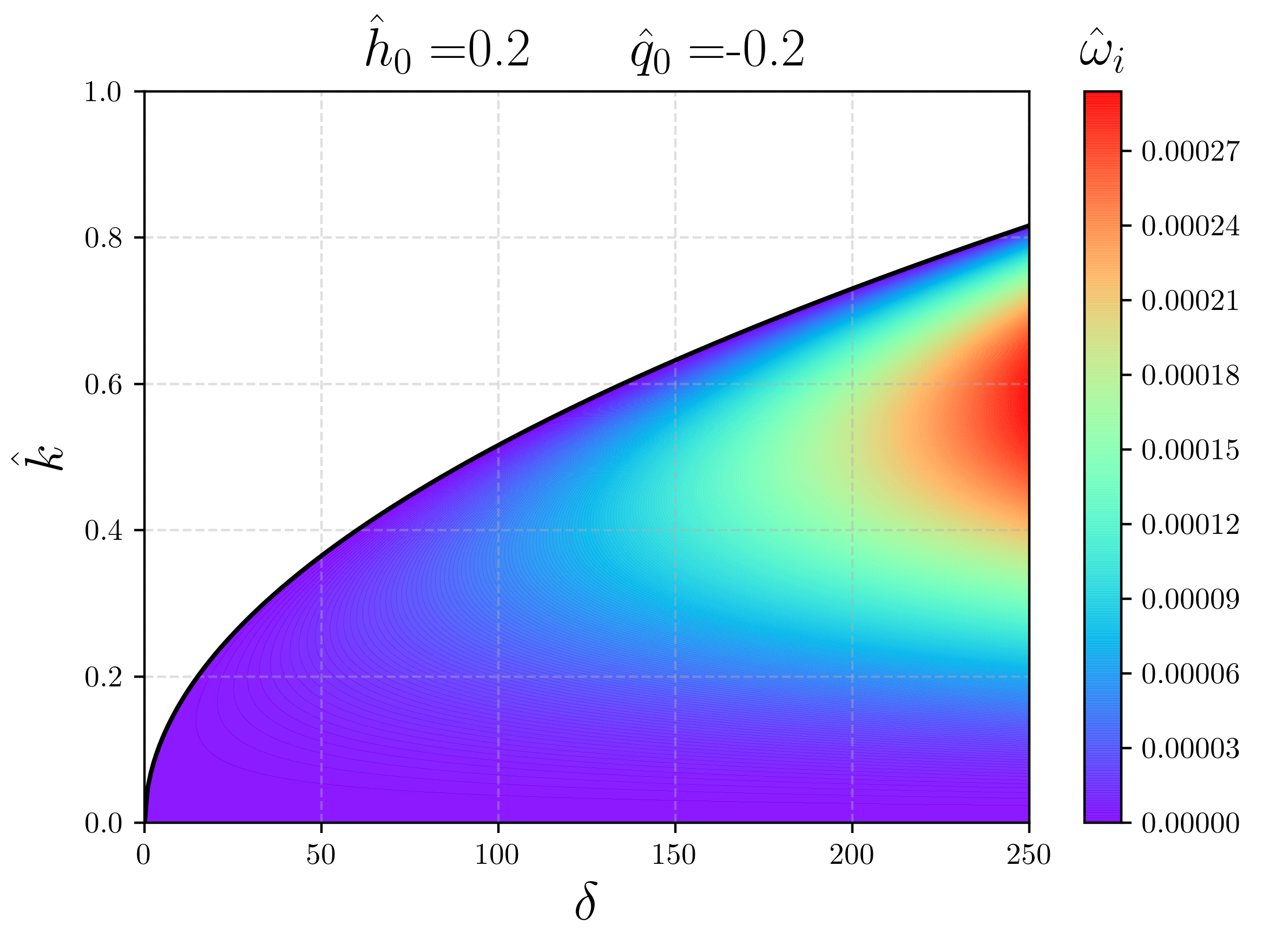}}
		
		\subfloat[Case for $\hat{h}_0=0.3$.]{\includegraphics[width=0.98\linewidth]{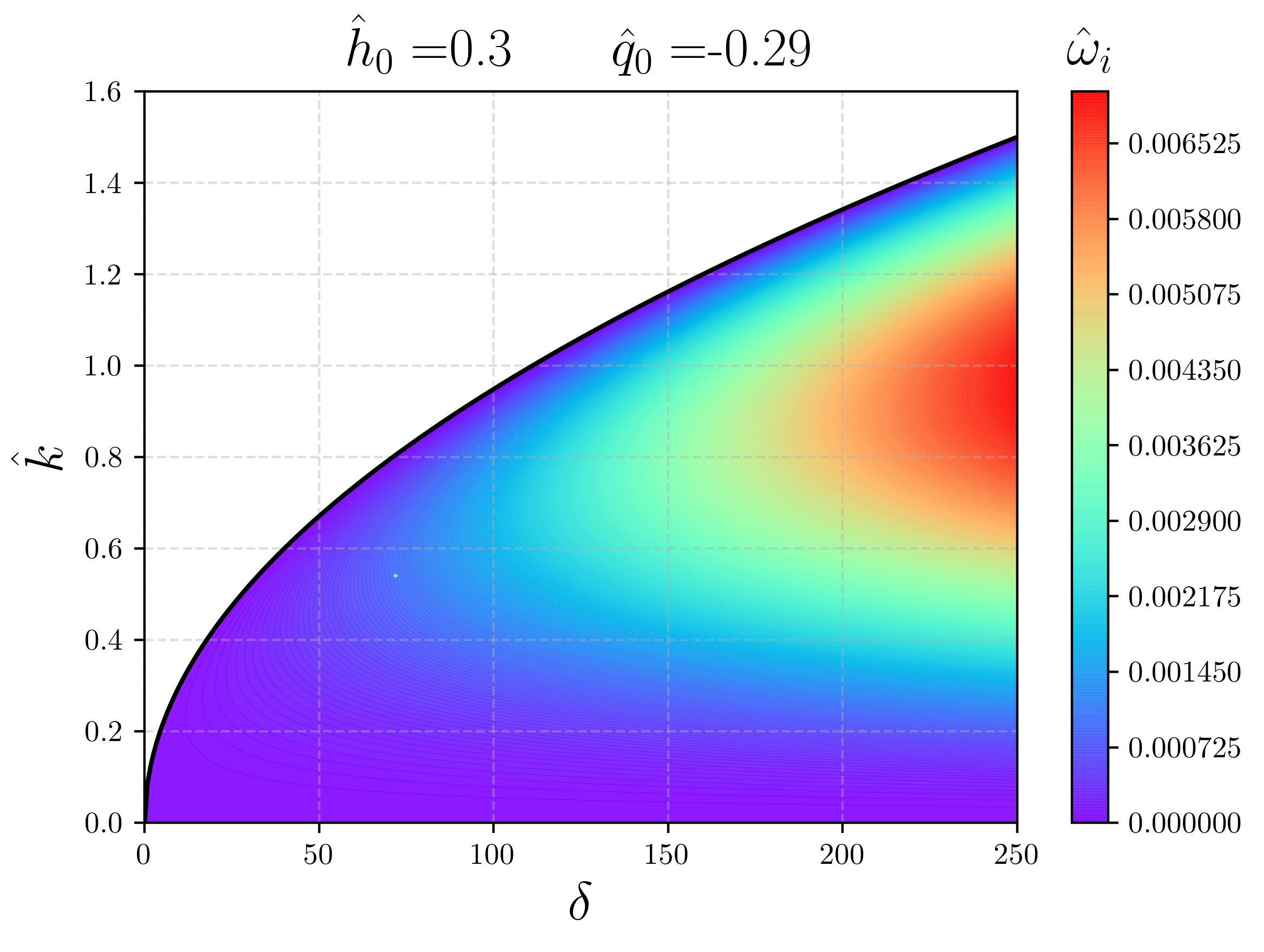}}
		\caption{Color map of $\log(\hat{\omega}_i)$ with the natural stability curve (black thick lines) for the MS case for $\hat{h}_0=0.1,0.2,0.3$. 
			The conditions with $\delta_1=76$ and $\delta_2=153$ analyzed in Sec. \ref{sec6p2} are located in the white zone further above these stability curves.
		}
		\label{fig:omega:MS}
	\end{figure}

	\begin{figure}[htbp]
		\centering
		\includegraphics[width=\linewidth]{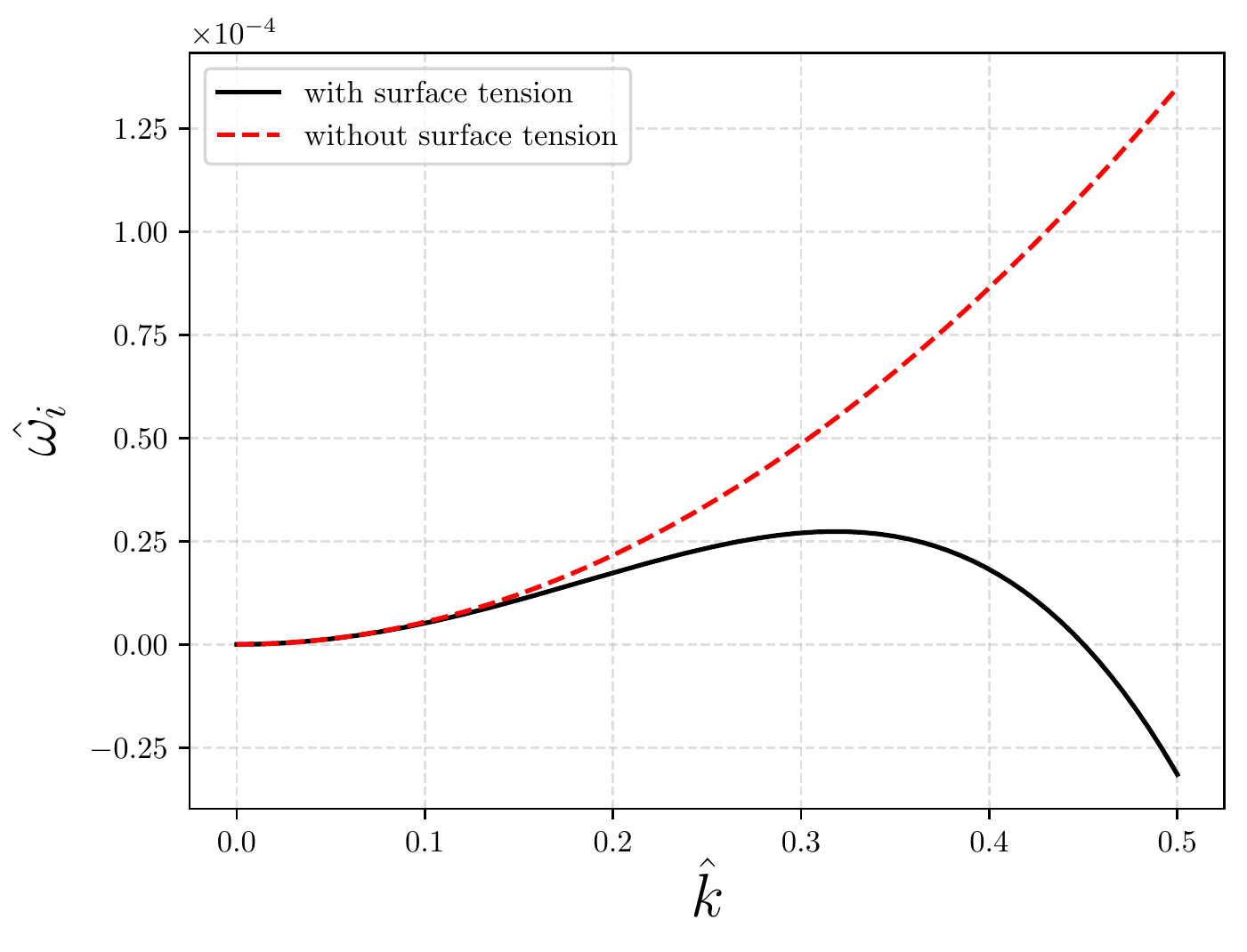}
		\caption{Plot of the imaginary part in the dispersion relation $\hat{\omega}_i(\hat{k})$ with and without surface tension for the case with $\delta=76$ and $\hat{h}_0=0.2$.}
		\label{fig:omega_k}
	\end{figure}

	The same plots are then produced for the MS problem, and shown in Fig. \ref{fig:omega:MS} for $\hat{h}_0=0.1,0.2,0.3$. We recall that in the MS problem one has $\alpha=-1$, hence $\hat{q}_0=\hat{h}_0^3/3- \hat{h}_0$. The location of the conditions analyzed in the nonlinear setting in section \ref{sec6p2} are not show since these are far away from the neutral curve. We have $k$ in the range $15-17$ for the case $\delta=76$ and in the range $21-23$ for the case $\delta=153$. The critical wave-number at these Reynolds number (see eq. 19b) is one order of magnitude lower than the one in the investigated conditions. Therefore, according to the linear stability analysis, all the investigated points should be in stable conditions.
	
	The region of maximum amplification is located at much larger $\delta$ compared to the FF and the region of largest amplification moves towards lower $\delta$.

	\subsection{Nonlinear analysis of 2-D waves}
	\label{sec6p2}
	
	We here move to the nonlinear analysis of the test cases introduced in Section \ref{sec5p3}. The perturbations are not infinitesimally small and the governing equations are not linearized. We recall that in this numerical investigation we do not include the contribution of the surface tension. Therefore, the dynamics of the liquid film is lacking the stabilizing effect: while all the investigated test cases are linearly stable if surface tension is included (cf. Fig. \ref{fig:omega:MS}), these are (linearly) unstable in absence of surface tension. We further illustrate this in Fig. \ref{fig:omega_k}, which shows the imaginary part of the dispersion relation $\hat{\omega}_i (\hat{k})$ with and without surface tension for the case with $\delta=76$ and $\hat{h}_0=0.2$. We recall, from Eq. \eqref{eq:19a_disp_rel}, that the dispersion relation is linear regardless of the surface tension and hence waves are non-dispersive. For the illustrated case, $\partial_{\hat{k}}\hat{\omega}_r=-0.96$, i.e. waves move approximately at the substrate speed.
	
	
	\begin{figure}[htbp]
		\centering
		\subfloat[Evolution of nonlinear waves at $\hat{h}_0=0.2$ and $\delta = 76$.]{\includegraphics[width=\linewidth]{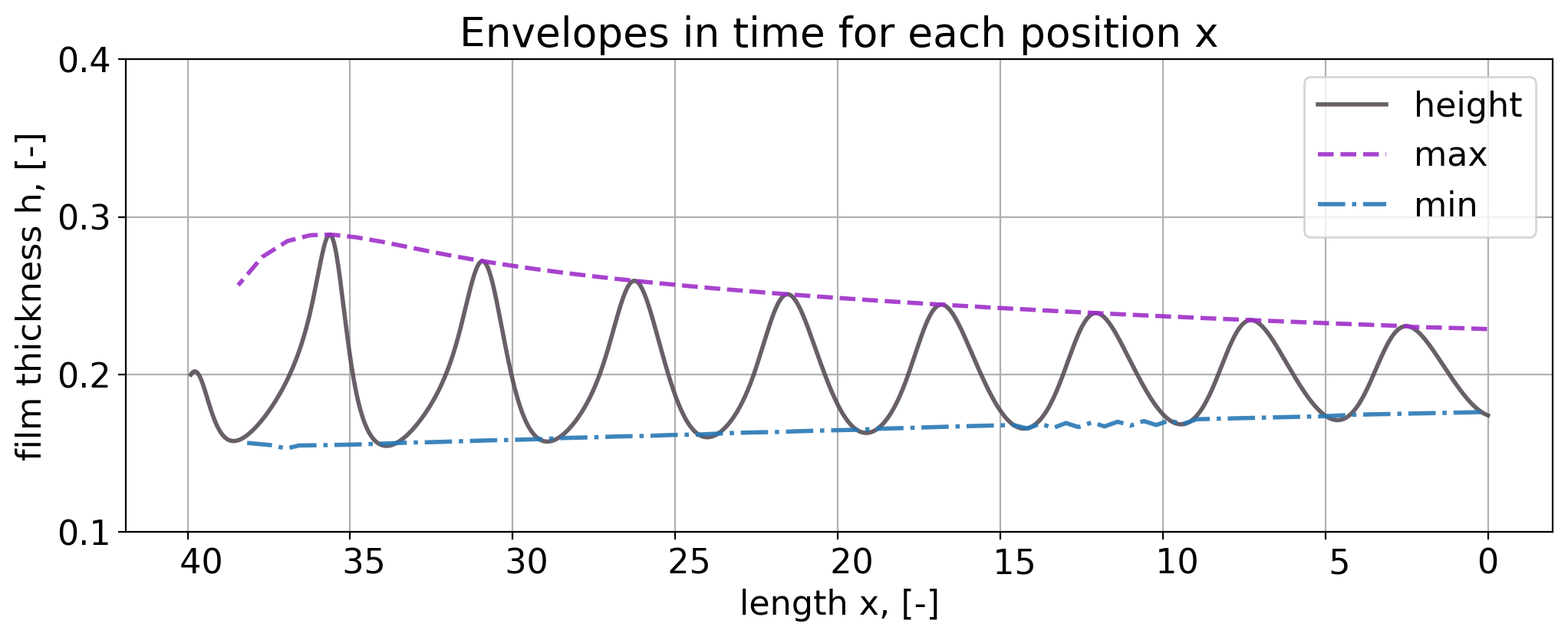}\label{fig:water_env}}
		
		\subfloat[Evolution of nonlinear waves at $\hat{h}_0=0.2$ and $\delta = 153$.]{\includegraphics[width=\linewidth]{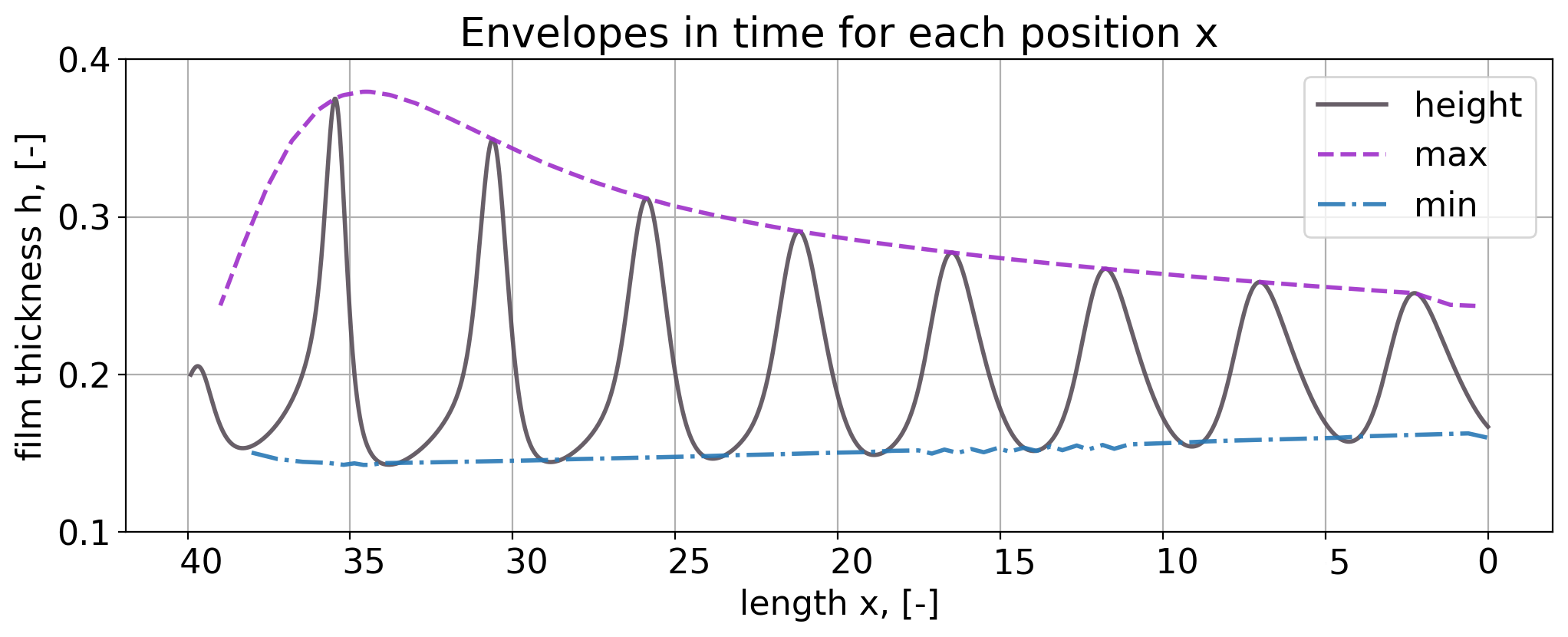}\label{fig:zinc_env}}
		
		\caption{Two-dimensional waves evolving without surface tension on a liquid film on a moving substrate with $\hat{h}_0 = 0.2$. The perturbation frequency is $\hat{f}=0.2$ ($\hat{\omega}_r=1.26$), with amplitude $\hat{q}_A = 0.2$. The substrate motion is from left to right; gravity is from right to left. The inlet is at $\hat{x}=40$.}
		\label{fig:envs}
	\end{figure}
	
	Moving to the nonlinear analysis in absence of surface tension, Fig. \ref{fig:envs} shows an instantaneous of the film thickness profile, together with the maxima and minima envelopes computed in the time for each position and considering $\hat{h}_0=0.2$ and $\delta_1=76$ (top) and $\delta_2=153$ (bottom). Despite the large disturbance ($\hat{q}_A=0.2$ in \eqref{eq:pert}) at the inlet ($\hat{x}=40$) and despite the initial growth of the waves, the flow remains convectively stable: the wave amplitude decreases in the stream-wise direction. These waves are nonlinear, with a steep front and a long tail similar to the waves observed in the FF problem (Fig. \ref{fig:doro}) in the first portion of the domain. However, the interplay of inertia, gravity and viscosity is different in the MS and FF cases. Contrary to the FF problem, the relative velocity between the flow and the wall \emph{decreases} at larger thicknesses if $\hat{h}_0\ll 1$. This implies that, contrary to the FF problem, the crest of the wave is \emph{slower} than the substrate film on which they travel. Therefore, waves tend to level out as they flow and this tendency is more pronounced for larger waves.
	
\begin{figure}[htbp]
		\centering
		
		\subfloat[Decay amplitudes for a reduced Reynolds number $\delta_1 = 76$.]{\includegraphics[width=0.9\linewidth]{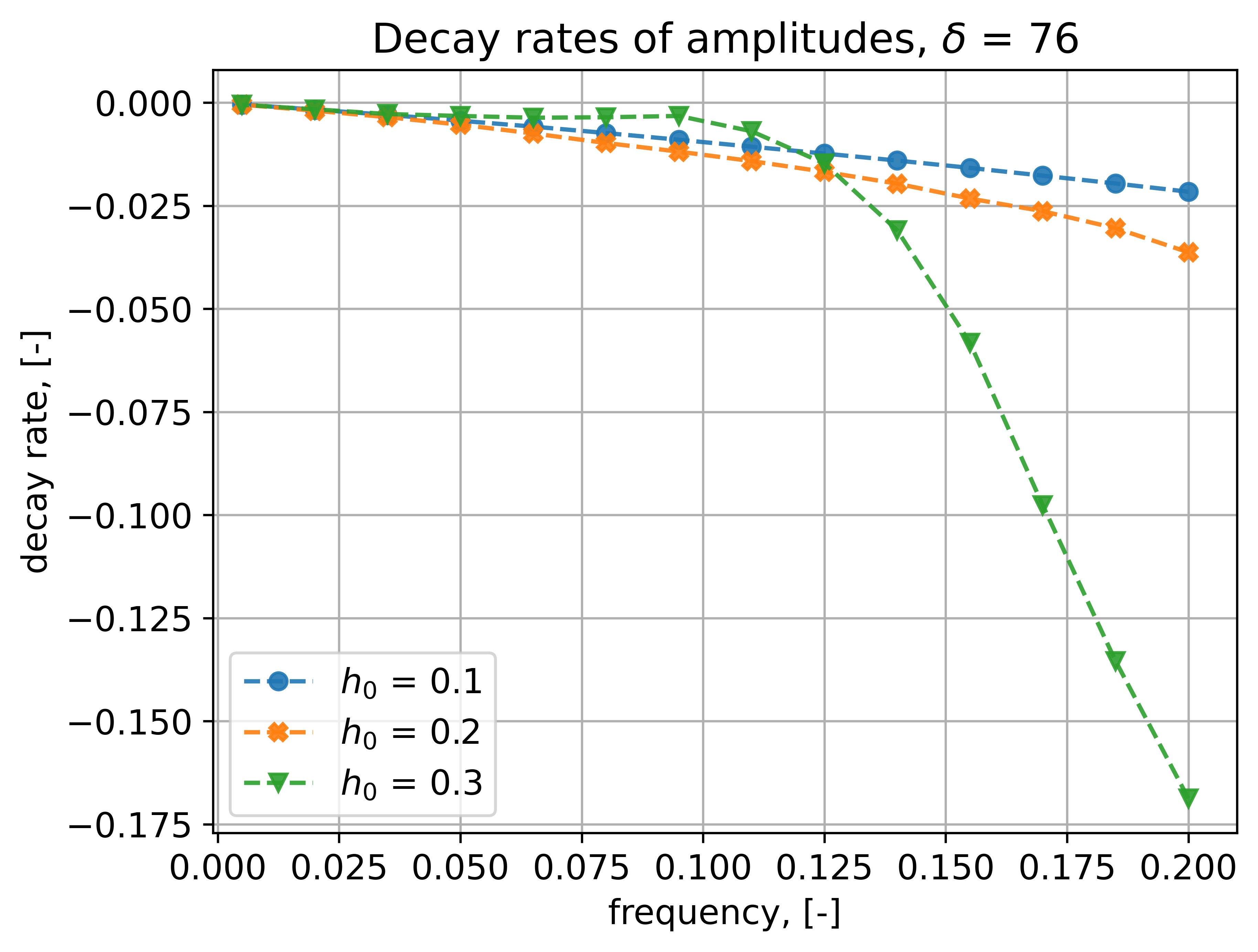}}
		
		\subfloat[Decay amplitudes for a reduced Reynolds number $\delta_2 = 153$.]{\includegraphics[width=0.9\linewidth]{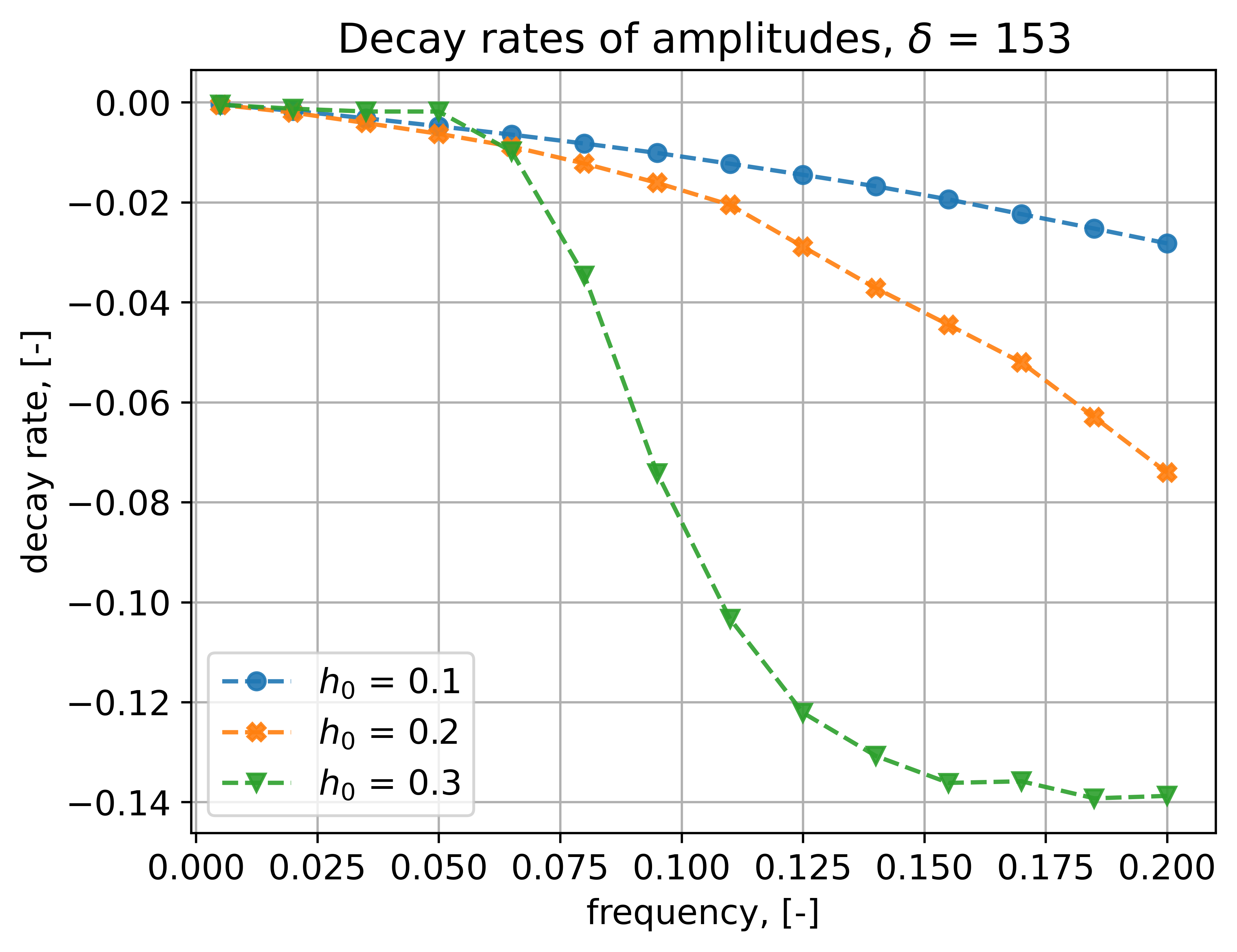}}
		
		\caption{Dependency of the water wave amplitude decay rates on the perturbation frequencies for three different initial heights $h_0$ and two different reduced Reynolds numbers $\delta_1 = 76$ and $\delta_2 = 153$.}
		\label{fig:decays}
	\end{figure}

	The initial growth produced in $30 < \hat{x} < 40$ 
	is most probably due to the mechanism through which the perturbations are injected (see Section \ref{sec5p2}). The boundary condition for injecting the perturbation simulate a manifold and the flow needs a certain distance to adjust back to the equations governing the film thickness and flow rate (see Section \ref{sec3}). Once this occurs, a clear decay of the wave amplitude is observed in all investigated configurations.
	We can thus fit an exponential decay $h_M e^{-\beta \hat{x}}$ to the maximum thickness temporal envelope and extract a spatial decay rate $\beta$ for each of the $84$ investigated simulations. The results are shown in Fig. \ref{fig:decays} for $\delta_1=76$ (top) and $\delta_2=153$ (bottom) over the range of perturbation frequencies ($\hat{f}\in[0.005,0.2]$) and for three thicknesses ($\hat{h}_0=0.1,0.2,0.3$). 
	
	While one would expect the decay rate of the waves to become stronger at larger frequencies, somewhat less expected is the fact that thicker films (within the investigated cases) lead to more substantial damping. In line with the damping mechanism previously described (shown in Fig. \ref{fig:envs}), thicker films result in larger waves at a short distance from the inlet, and larger waves are slower and thus characterized by stronger damping. This mechanism is promoted by the nonlinearities of the problem and is not captured in the linear stability framework (which predicts unstable waves in the absence of surface tension).
	
	\begin{figure}[htbp]
		\centering
		
		\subfloat[spatiotemporal maps of the evolution of 2-D waves in a liquid film over a moving substrate.]{\includegraphics[width=0.85\linewidth]{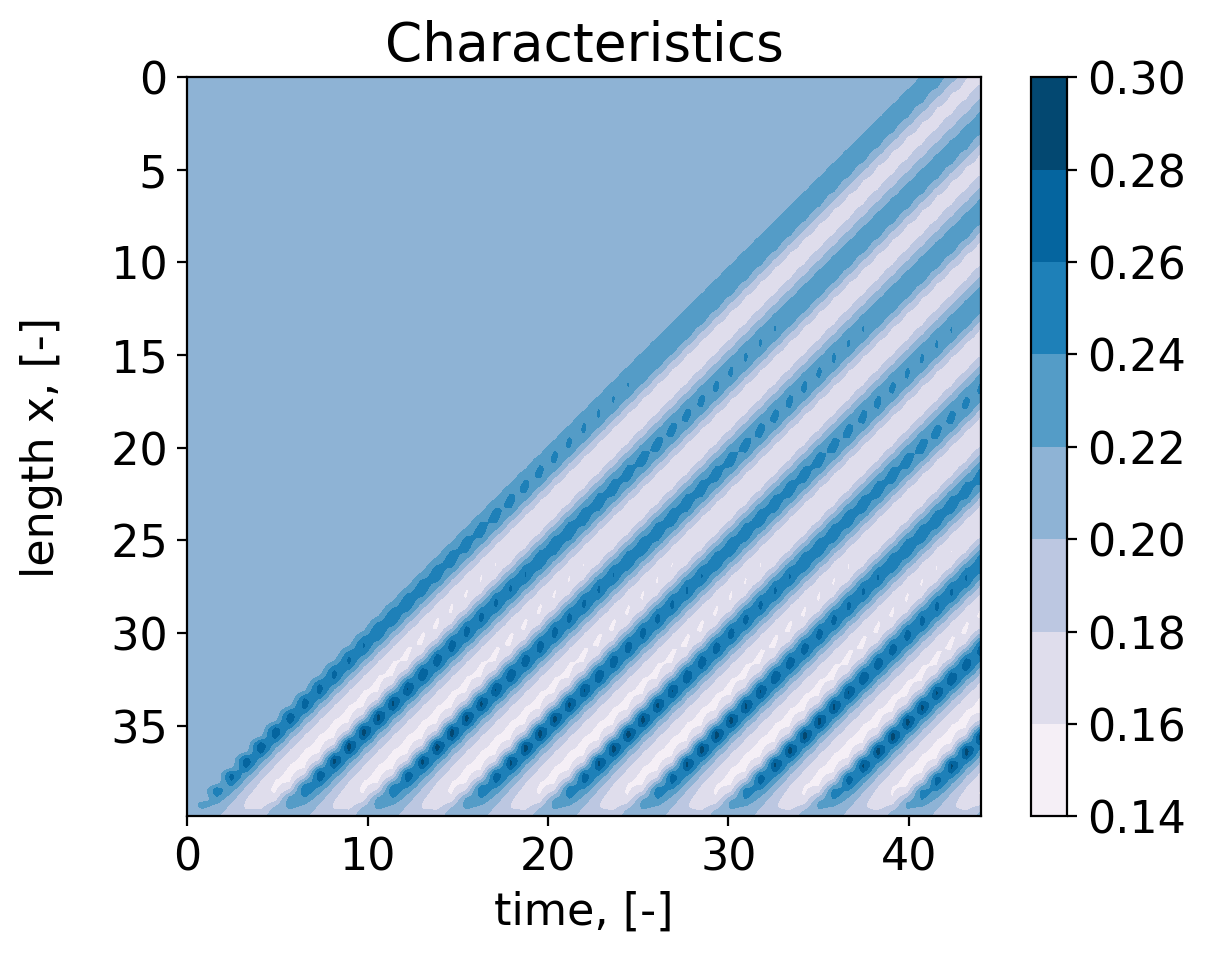}\label{fig:char}}
		
		\subfloat[Neither wave merging, nor frequency cross-talk is observed.]{\includegraphics[width=0.85\linewidth]{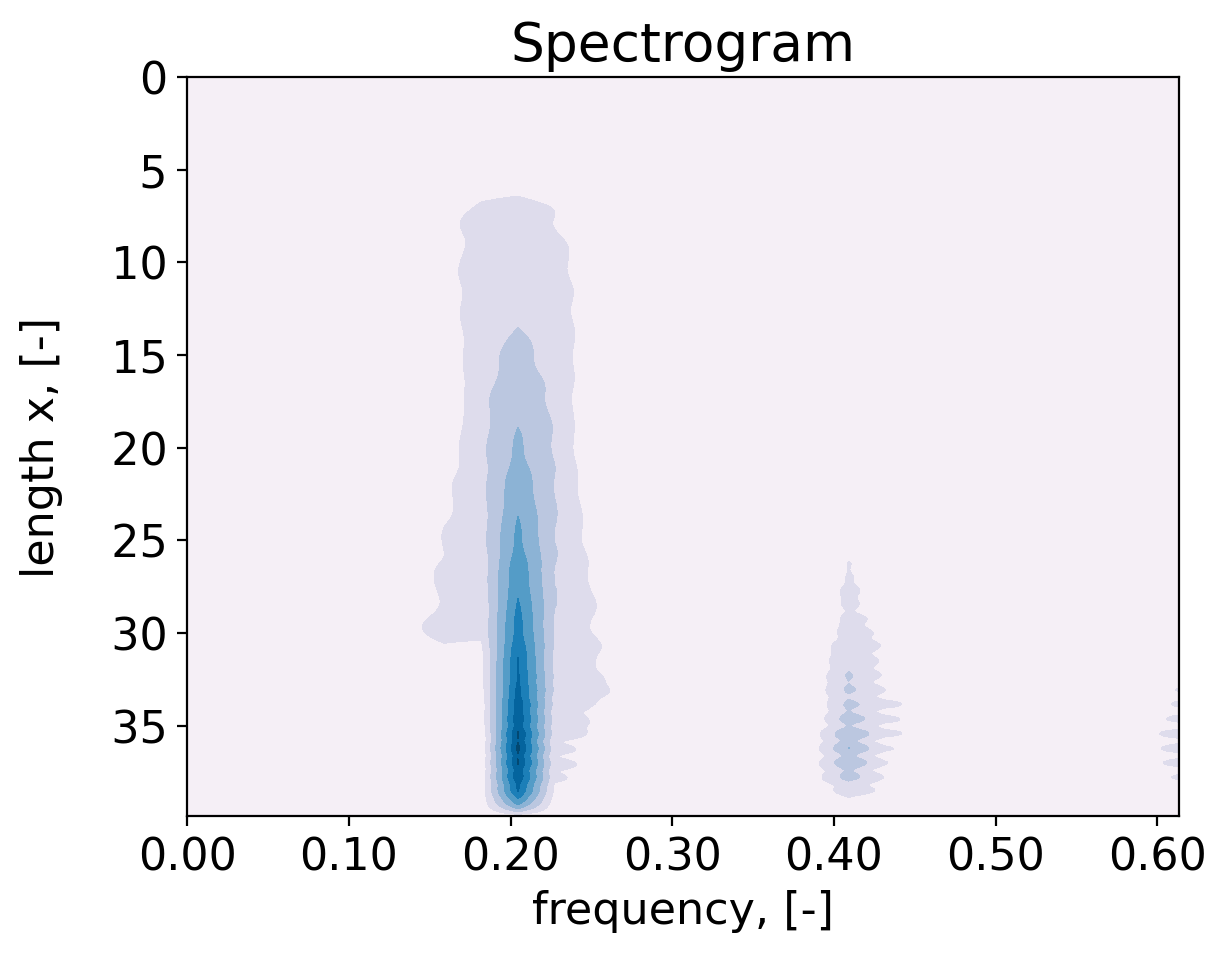}\label{fig:spect}}
		
		\caption{The operating conditions in these plots are the same as for the water wave in Fig. \ref{fig:envs} (top) with $\hat{h}_0 = 0.2, \delta = 76, \hat{f} = 0.2$.}
		\label{fig:char_spect}
	\end{figure}

	Finally, we highlight that during the downstream propagation, the phase velocity is approximately constant and equal to $\approx -1$, i.e., substrate velocity. This can be seen from Fig. \ref{fig:char}, which shows the spatiotemporal evolution of the waves for a case with $\hat{h}_0=0.2$, $\delta=76$ and $\hat{f}=0.3$. The characteristic lines are straight, showing no appreciable acceleration or deceleration during their evolution, and their velocity is line with the linear stability analysis. Fig. \ref{fig:spect} shows the evolution of the frequency content of the thickness evolution at various locations. The nonlinearities do not produce wave-merging mechanisms within the investigated test cases: the same frequency is propagated downstream, and higher harmonics (linked to the initial stiff front of the waves) gently vanishes as the wave amplitude decreases and their profile becomes more harmonic.

	\subsection{Analysis of 3-D waves}
	\label{sec6p3}
	
\begin{figure}[htbp]
		\centering
		
		\subfloat[Dimensionless liquid film height of a 3-D wave generated by perturbations.]{\includegraphics[width=0.9\linewidth]{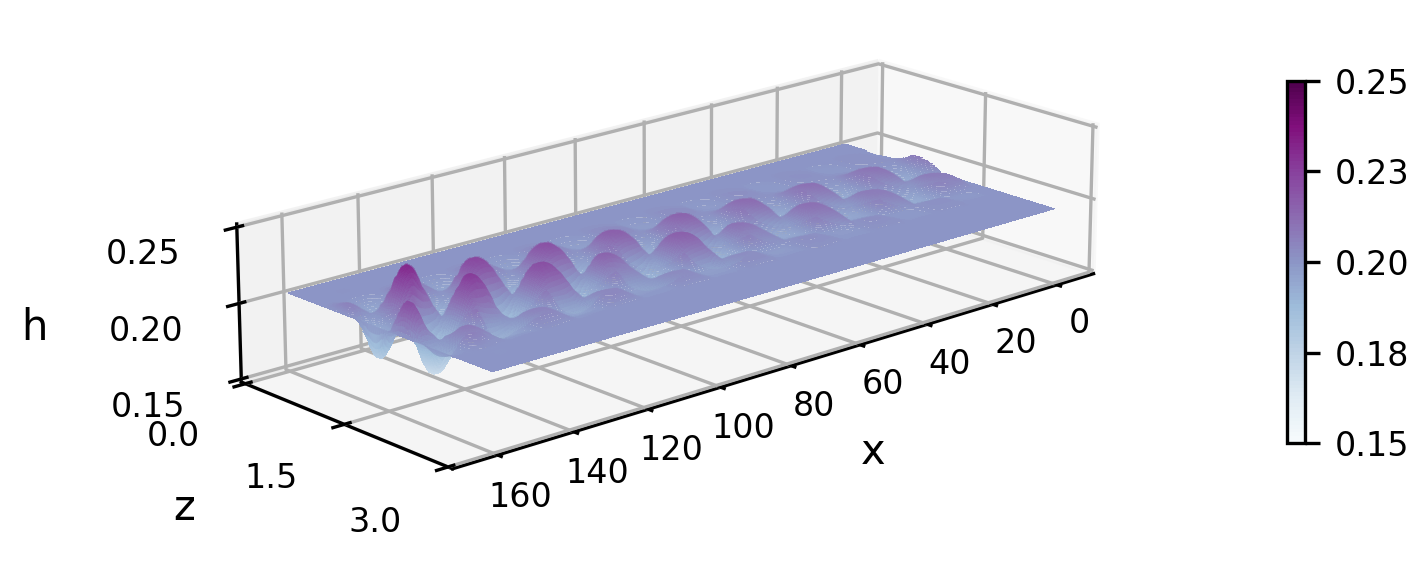}
			\label{fig:3dsurf}}
		
		\subfloat[Contour plot of the height of the 3-D wave.]{\includegraphics[width=0.9\linewidth]{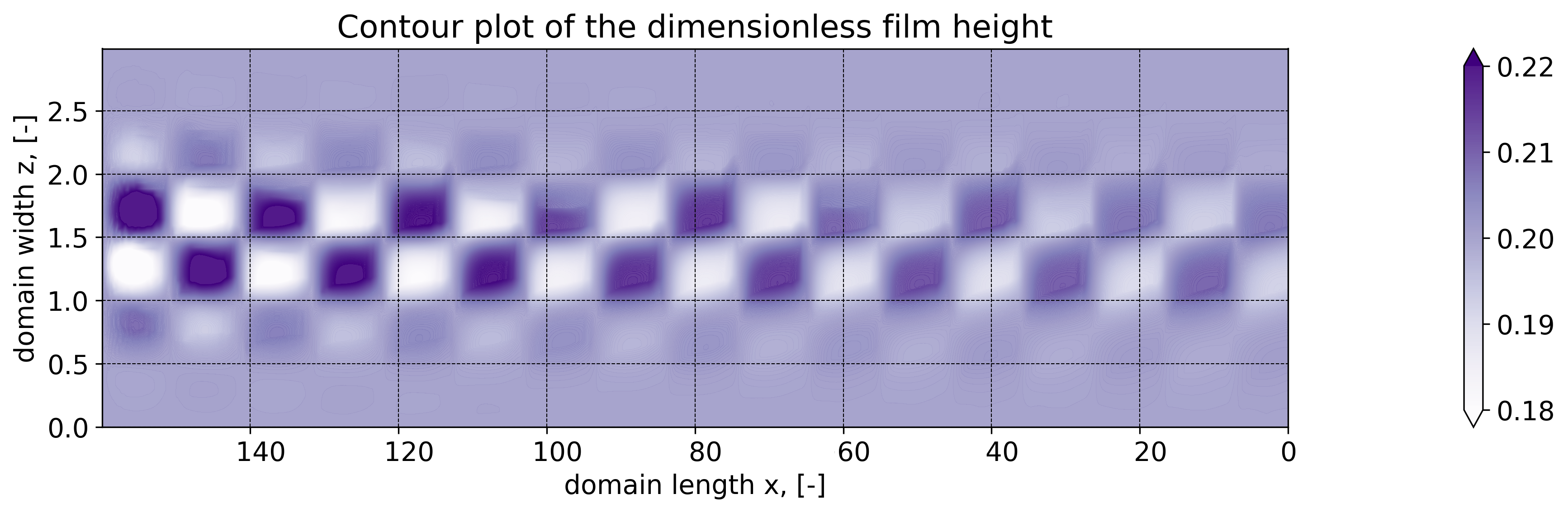}
			\label{fig:3dcont}}
		
		\caption{Dimensionless liquid film height (top) and its contour plot (bottom) for 3-D waves generated by the flow rate perturbation of $\hat{q}_x$ in Eq. \eqref{eq:pert_qx}. The substrate moves toward $\hat{x}\rightarrow-\infty$ while gravity is directed toward $\hat{x}\rightarrow \infty$ (see Fig. \ref{fig:jw3d}).
			The disturbances are introduced on the left, at $\hat{x}=160$.
			The dimensionless perturbation frequency is $\hat{f} = 0.05$, with initial film thickness $\hat{h}_0 = 0.2$ and $\delta = 76$.}
		\label{fig:surf}	
	\end{figure}

	We conclude this investigation with the analysis of a three-dimensional test case, with inlet flow rate $\hat{q}_x$ prescribed as in \eqref{eq:pert_qx}, and inlet flow rate $\hat{q}_z = 0$. The main interest was to analyze if and how three-dimensional perturbations grow in the spanwise direction $\hat{z}$ (see also Fig. \ref{fig:jw3d}). A snapshot of the liquid film surface for this test case is shown in Fig. \ref{fig:3dsurf}, while Fig. \ref{fig:3dcont} shows the film thickness contour plot. 
	
	The results show that the motion of the substrate dominates the direction of the propagation and the region of influence of the disturbance is particularly narrow in the $\hat{z}$ direction.
	This is in contrast to what happens, for example, in the waves of shallow or deep waters where a perturbation in relative motion with respect to the substrate (e.g. a ship) produces a V-shaped wake envelope in case of shallow (non-dispersive) waves and the well-known Kelvin wedge in the case of deep (dispersive) waves \citet{Fitzpatrick2019,Lighthill2007}. Although we leave the analysis of more general kinds of 3-D disturbances to future works, these results highlight the distinctive role of nonlinearities in the MS problem compared to other cases of nonlinear waves in fluid dynamics.

	\section{Conclusions}
	\label{sec7}
	
	We presented integral boundary layer models to describe the evolution of interface waves in liquid films dragged along upward-moving walls (MS problem) and compared the problem with the well-known case of liquid films falling along fixed walls (FF problem). We introduced 
	a dimensionless 3-D integral boundary layer model that extends classic models for falling liquid films to account for substrate motion, external pressure gradient and interface shear stress. These extensions allow for modeling the jet wiping process in hot-dip galvanization.
	
	The interest in integral models is twofold: they allow for performing computationally inexpensive numerical simulations of liquid film flows and enable analytical insights on their stability. We analyzed the stability of the MS problem in a linear and nonlinear setting. For the linear setting, focusing on 2-D disturbances, we derived dispersion relations and neutral curves, and compared the results with those of the FF problem. For the nonlinear setting, we analyzed the flow response to 2-D disturbances of various frequencies, Reynolds number and baseline film thickness numerically. These conditions are relevant to industrial applications.
	
	In the linear stability analysis, the dispersion relation shows the usual stabilizing effect of surface tension. It was shown that the critical wave number is smaller than in the FF problem if $\hat{h}_0<1$. In the nonlinear stability analysis, a nonlinear stabilizing mechanism was identified even in the absence of surface tension. The wave amplitude decays because the interface velocity is lower than the substrate, and this leveling effect is larger for thicker films. Finally, we presented a numerical test case with 3-D disturbances. This showed no disturbance growth in the span-wise direction.
	
	Future work will extend the current analysis to the evolution of other kind of 3-D disturbances, other integral models (e.g., the Weighted IBL formulation) and the Orr-Sommerfeld problem from the full Navier-Stokes equations.

	\begin{acknowledgements} 
		This work is supported by ArcelorMittal in the framework of the 'Ondule VII' project. Fabio Pino is supported by a FRIA grant from FNRS. Benoit Scheid thanks the F.R.S.-FNRS for financial support. 
		
		The solver and post-processing scripts have been developed using free and open-source software: we thank the community for building powerful tools from which everyone can benefit. 
		
		Finally, we thank David Barreiro from University of A Coruña for providing valuable feedback and correcting a mistake in the derivation of Eqs. \eqref{eq:17a} - \eqref{eq:17b}.
		
	\end{acknowledgements}

	\appendix
	
	\section{The full (dimensional) problem}
	\label{app:NS}
	The goal is to obtain an integral model for the 3-D liquid film on a moving substrate, since this approach reduces the number of independent variables and the dimension of the domain. This leads to a significantly lower computational cost when performing numerical analysis as shown in \citet{mendez_jfm}. The developed finite volume solver for the purposes of this research is a 3-D extension of the one in \citet{mendez_jfm}.
	
	The starting point for the formulation of the integral model is the Navier-Stokes equations for a divergence-free Newtonian liquid (\citet{graebel2007}):
	\begin{eqnarray}
		\frac{\partial \vec{v}}{\partial t} + \vec{v} \cdot \nabla \vec{v} = - \frac{1}{\rho} \nabla p + \nu \nabla^2 \vec{v} + \vec{f_v},
		\label{eq:Navier-Stokes}
	\end{eqnarray}
	where $\rho$ is the density of the liquid, $\nu$ is its kinematic viscosity, $\vec{v} = (u,v,w)$ is the velocity field, and $p$ is the pressure field. 
	
	The continuity equation for a divergence-free fluid in a Cartesian coordinate system states 
	\begin{eqnarray}
		\label{eq:cont}
		\nabla \cdot \vec{v} = 0.
	\end{eqnarray}
	The kinematic boundary conditions at the wall consist of the non-slip and non-permeability conditions for this problem:
	\begin{eqnarray}
		\vec{v} \big|_{y=0} = (-U_p, 0, 0),
		\label{eq:bc-kinematic-wall}
	\end{eqnarray}
	where $U_p$ is the speed of the substrate. The continuity of the interface $h(x,z,t)$ is ensured by the kinematic boundary condition
	\begin{eqnarray}
		\frac{\partial}{\partial t}(y-h(x,z,t)) + \vec{v} \cdot \nabla (y-h(x,z,t)) = 0 \nonumber
	\end{eqnarray}
	at $y = h(x,z,t)$ rewritten as
	\begin{eqnarray}
		v = \partial_t h + u \partial_x h + w \partial_z h.
		\label{eq:bc-kinematic-interface}
	\end{eqnarray}
	
	The dynamic boundary conditions at the interface represents the force balance in a local Cartesian reference frame along the free surface normal $\vec{n}$, and the streamwise and spanwise tangential directions, denoted by $\vec{t}_x$ and $\vec{t}_z$, respectively.
	These vectors have the following components:
	\begin{subequations}
		\label{vectors}
		\begin{eqnarray}
			\vec{n} &=& \frac{1}{|\vec{n}|} \grad (y-h(x,z,t)) \\ &=& \frac{1}{|\vec{n}|} (-\partial_x h(x,z,t), 1, -\partial_z h(x,z,t)), \\
			\vec{t}_x &=& \frac{1}{|\vec{t}_x|} (1, \partial_x h(x,z,t), 0), \\
			\vec{t}_z &=& \frac{1}{|\vec{t}_z|} (0, \partial_z h(x,z,t), 1).
		\end{eqnarray}
	\end{subequations}
	To denote quantities referring to the gas, the subscript $g$ is introduced to the variables notations. 
	At the interface $y = h(x,z,t)$, the force balances along these directions are:
	\begin{subequations}
		\label{eq:balances}
		\begin{eqnarray}
			\vec{n} \cdot \hat{T} \cdot \vec{n} - \vec{n} \cdot \hat{T}_g \cdot \vec{n} &=& \sigma \divg \vec{n}, \\
			\vec{n} \cdot \hat{T} \cdot \vec{t}_x - \vec{n} \cdot \hat{T}_g \cdot \vec{t}_x &=& 0, \\
			\vec{n} \cdot \hat{T} \cdot \vec{t}_z - \vec{n} \cdot \hat{T}_g \cdot \vec{t}_z &=& 0.
		\end{eqnarray}
	\end{subequations}
	The stress tensor for the two fluids is:
	\begin{eqnarray}
		\label{eq:stress_tensor}
		\hat{T} = -p \hat{I} + 2 \mu \hat{E},
	\end{eqnarray}
	where $\hat{I}$ is the identity tensor, and $\hat{E}$ is the strain of rate tensor that is \begin{eqnarray}
		\hat{E} &=& \frac{1}{2} ( \grad \vec{v} + \grad \vec{v}^T ).
	\end{eqnarray}
	It is worth remarking that $\vec{n} \cdot (-p \hat{I}) \cdot \vec{n} = -p$ and $\vec{n} \cdot (-p \hat{I}) \cdot \vec{\tau}_{x,z} = 0$ both for the liquid and the gas.
	Substituting the stress tensor definition \eqref{eq:stress_tensor} into equations \eqref{eq:balances} yields: 
	\begin{eqnarray}
		-p + \vec{n} \cdot (2\mu \hat{E}) \cdot \vec{n} + p_g - \vec{n} \cdot (2\mu \hat{E}_g) \cdot \vec{n} &=& \sigma \divg \vec{n}, \nonumber\\
		\vec{n} \cdot (2\mu \hat{E}) \cdot \vec{t}_x - \vec{n} \cdot (2\mu \hat{E}_g) \cdot \vec{t}_x &=& 0, \nonumber \\
		\vec{n} \cdot (2\mu \hat{E}) \cdot \vec{t}_z - \vec{n} \cdot (2\mu \hat{E}_g) \cdot \vec{t}_z &=& 0. \nonumber
	\end{eqnarray}
	By introducing the following notations
	\begin{eqnarray}
		p_g - \vec{n} \cdot (2\mu \hat{E}_g) \cdot \vec{n} &=& p_g (x,z,t), \nonumber\\
		\vec{n} \cdot (2\mu \hat{E}_g) \cdot \vec{t}_x &=& \tau_{g,x}(x,z,t), \nonumber \\
		\vec{n} \cdot (2\mu \hat{E}_g) \cdot \vec{t}_z &=& \tau_{g,z}(x,z,t), \nonumber
	\end{eqnarray}
	we reach the three scalar equations representing the projections of the force balance along the normal and tangential vectors:
	\begin{subequations}
		\label{eq:forces-balances}
		\begin{eqnarray}
			-p + \vec{n} \cdot (2\mu \hat{E} ) \cdot \vec{n} + p_g(x,z,t) &=& \sigma \divg \vec{n}, \\
			\vec{n} \cdot (2\mu \hat{E}) \cdot \vec{t}_x - \tau_{g,x}(x,z,t) &=& 0, \\
			\vec{n} \cdot (2\mu \hat{E} ) \cdot \vec{t}_z - \tau_{g,z}(x,z,t) &=& 0.
		\end{eqnarray}
	\end{subequations}
	The computation of the expressions $\vec{n} \cdot (2\mu \hat{E} ) \cdot \vec{n}, \vec{n} \cdot (2\mu \hat{E} ) \cdot \vec{t}_x$ and $\vec{n} \cdot (2\mu \hat{E}) \cdot \vec{t}_z$ in a 3-D Cartesian coordinate system is a lengthy procedure and its details are therefore omitted in this paper. It results in the dynamic boundary conditions that represent the force balance at the interface with projections along $\vec{n}, \vec{t}_x$, and $\vec{t}_z$, which is shown in Section \ref{sec3} in a first-order boundary layer approximation followed by the integration along the wall-normal $y$-axis.
	
	The 3-D liquid film on a moving substrate is therefore represented by the dimensional equations \eqref{eq:Navier-Stokes} and \eqref{eq:cont} for an incompressible Newtonian fluid with boundary conditions \eqref{eq:bc-kinematic-wall}, \eqref{eq:bc-kinematic-interface}, and \eqref{eq:forces-balances}.
	The scaling procedure of these equations and boundary conditions with appropriate reference quantities is described in Section \ref{sec2}.

	\section{Solver and numerical schemes} 
	\label{app:num}
	
	The numerical solution of the hyperbolic problem \eqref{eq:model_gen}, a blending between the Lax-Wendroff and the Lax-Friedrichs schemes is applied. This blending is achieved with flux limiter functions. They are used to help preventing oscillations near sharp changes in the solution.
	
	Let the discretized state vector be denoted by $\vec{U}$.
	In order to evaluate the state at the next time step, estimations of the quantities at mid-points in space and time are necessary:
	\begin{subequations}
		\begin{eqnarray}
			\vec{U}_{i+\frac{1}{2}, j}^{n+\frac{1}{2}} = \frac{1}{2} \big(\vec{U}_{i, j}^{n} + \vec{U}_{i+1, j}^{n}\big) - \frac{\Delta t}{2 \Delta x} \big( \vec{\mathcal{F}_{x,}}_{i+1,j}^{n} - \vec{\mathcal{F}_{x,}}_{i,j}^{n} \big), \\
			\vec{U}_{i, j+\frac{1}{2}}^{n+\frac{1}{2}} = \frac{1}{2} \big(\vec{U}_{i, j}^{n} + \vec{U}_{i, j+1}^{n}\big) - \frac{\Delta t}{2 \Delta z} \big( \vec{\mathcal{F}_{z,}}_{i,j+1}^{n} - \vec{\mathcal{F}_{z,}}_{i,j}^{n} \big).
		\end{eqnarray}
	\end{subequations}
	
	The fluxes can be represented by low and high-resolution schemes and a flux limiter can switch between these schemes depending on the gradients of the solutions.
	Values of the fluxes at the half steps are evaluated in the following way.
	The high-resolution fluxes are:
	\begin{subequations}
		\begin{eqnarray}
			\vec{\mathcal{F}_{x,}}_{i + \frac{1}{2}, j}^{\text{high}} &=&  \vec{\mathcal{F}_{x}}(\vec{U}_{i + \frac{1}{2}, j}^{n + \frac{1}{2}}), \\
			\vec{\mathcal{F}_{z,}}_{i, j + \frac{1}{2}}^{\text{high}} &=&  \vec{\mathcal{F}_{z}}(\vec{U}_{i, j + \frac{1}{2}}^{n + \frac{1}{2}}).
		\end{eqnarray}
	\end{subequations}
	The low-resolution fluxes are:
	\begin{subequations}
		\begin{eqnarray}
			\vec{\mathcal{F}_{x,}}_{i + \frac{1}{2}, j}^{\text{low}} = \vec{\mathcal{F}_{x,}}(\vec{U}_{i + 1, j}^{n}) + \frac{\Delta t}{2 \Delta x} \Big( \vec{U}_{i+\frac{1}{2}, j}^{n+\frac{1}{2}} - \vec{U}_{i+1, j}^{n} \Big), \\
			\vec{\mathcal{F}_{z,}}_{i, j + \frac{1}{2}}^{\text{low}} = \vec{\mathcal{F}_{z,}}(\vec{U}_{i, j + 1}^{n}) + \frac{\Delta t}{2 \Delta z} \Big( \vec{U}_{i, j+\frac{1}{2}}^{n+\frac{1}{2}} - \vec{U}_{i, j+1}^{n} \Big).
		\end{eqnarray}
	\end{subequations}
	The blended fluxes are:
	\begin{subequations}
		\begin{eqnarray}
			\vec{F_{x,}}_{i + \frac{1}{2}, j} = \vec{\mathcal{F}_{x,}}_{i + \frac{1}{2}, j}^{\text{low}} - \phi_x (r_{x, i}) \Big( \vec{\mathcal{F}_{x,}}_{i + \frac{1}{2}, j}^{\text{low}} - \vec{\mathcal{F}_{x,}}_{i + \frac{1}{2}, j}^{\text{high}} \Big) \\
			\vec{F_{x,}}_{i - \frac{1}{2}, j} = \vec{\mathcal{F}_{x,}}_{i - \frac{1}{2}, j}^{\text{low}} - \phi_x (r_{x, i-1}) \Big( \vec{\mathcal{F}_{x,}}_{i - \frac{1}{2}, j}^{\text{low}} - \vec{\mathcal{F}_{x,}}_{i - \frac{1}{2}, j}^{\text{high}} \Big) \\
			\vec{F_{z,}}_{i, j + \frac{1}{2}} = \vec{\mathcal{F}_{z,}}_{i, j + \frac{1}{2}}^{\text{low}} - \phi_z (r_{z, i}) \Big( \vec{\mathcal{F}_{z,}}_{i, j + \frac{1}{2}}^{\text{low}} - \vec{\mathcal{F}_{z,}}_{i, j + \frac{1}{2}}^{\text{high}} \Big) \\
			\vec{F_{z,}}_{i, j - \frac{1}{2}} = \vec{\mathcal{F}_{z,}}_{i, j - \frac{1}{2}}^{\text{low}} - \phi_z (r_{z, i-1}) \Big( \vec{\mathcal{F}_{x,}}_{i, j - \frac{1}{2}}^{\text{low}} - \vec{\mathcal{F}_{z,}}_{i, j - \frac{1}{2}}^{\text{high}} \Big)
		\end{eqnarray}
	\end{subequations}
	where $\phi_x, \phi_z$ are the flux limiter functions in $x$- and $z$-directions, and $r$ represents the ratio of successive gradients on the mesh:
	\begin{eqnarray}
		r_i &=& \frac{\vec{U}_i - \vec{U}_{i-1}}{\vec{U}_{i+1} - \vec{U}_i}.
	\end{eqnarray}
	Finally, using an explicit scheme, the time stepping for the solution vector is:
	\begin{eqnarray}
		\label{eq:blended-scheme}
		\vec{U}_{i,j}^{n+1} = \quad \vec{U}_{i,j}^{n} &-& \frac{\Delta t}{\Delta x} \Big( \vec{F_{x,}}_{i + \frac{1}{2}, j} - \vec{F_{x,}}_{i - \frac{1}{2}, j} \Big) \nonumber \\
		&-& \frac{\Delta t}{\Delta z} \Big( \vec{F_{z,}}_{i, j + \frac{1}{2}} - \vec{F_{z,}}_{i, j - \frac{1}{2}} \Big) \nonumber \\
		&+& \Delta t \vec{S}_{i,j}^n.
	\end{eqnarray}
	
	Depending on the flux limiter functions, different schemes can be obtained from the blended scheme \eqref{eq:blended-scheme}. 
	More specifically, if all values of the flux limiters are set to $1$, then the solution is smooth and the fluxes are represented by a high-resolution scheme. Substituting all values of the flux limiter functions with $1$ retrieves the Lax-Wendroff scheme. Another example is the Lax-Friedrichs scheme which can be obtained if all flux limiter values are $0$, which means that a low-resolution approximation of the fluxes is needed. 
	
\begin{figure}[htbp]
		\centering
		\includegraphics[width=0.6\linewidth]{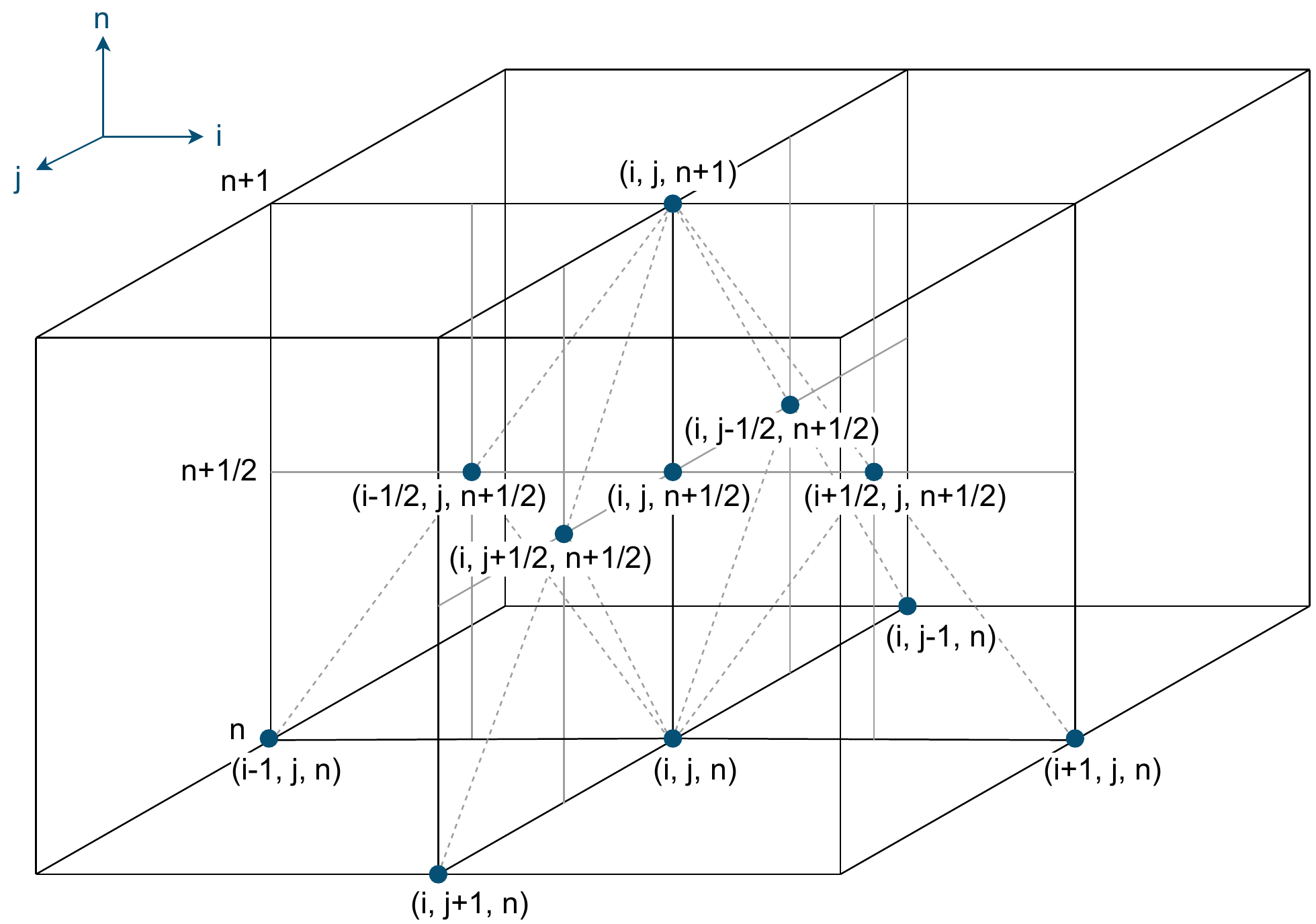}
		\caption{Stencil for the implemented numerical schemes. The spatial dimensions are represented by $i$ and $j$, and the time steps are indicated by $n$.}
		\label{fig:stencil}
	\end{figure}

	The preformed simulations in this work use also the \texttt{minmod} limiter functions which are of the following kind:
	\begin{eqnarray}
		\label{eq:minmod}
		&& \phi_{x} = \max[0, \min(1,r_x)], \quad
		\lim_{r_x \to \infty} \phi_{x} (r_x) = 1, 
		\\[0.5em]
		&& \phi_{z} = \max[0, \min(1,r_z)], \quad
		\lim_{r_z \to \infty} \phi_{z} (r_z) = 1. 
	\end{eqnarray}
	\\
	The numerical stencil for these two-dimensional schemes is presented in Fig. \ref{fig:stencil}. 
	The horizontal plane defined by $i$- and $j$-directions (the $xz$-domain) physically represents the liquid film height. The third dimension of the stencil $n$ represents the time.
	
	\bibliography{Ivanova_et_al_2023}
	
\end{document}